\documentclass[preprint,review,12pt]{elsarticle}

\usepackage{graphicx}
\usepackage{amssymb}
\usepackage{hyperref}
\usepackage{lineno}
\usepackage{float}
\usepackage{multirow}
\usepackage[table,xcdraw]{xcolor}
\usepackage[labelfont=bf]{caption}
\usepackage{todonotes}
\usepackage{multibib}
\newcites{New}{References used in SI}

\biboptions{comma,square, sort&compress}

\begin{document}

\begin{frontmatter}

\title{The impact of Kelvin probe force microscopy operation modes and environment on grain boundary band bending in perovskite and Cu(In,Ga)Se$_2$ solar cells}



\author{Evandro Martin Lanzoni}
\ead{evandro.lanzoni@uni.lu}

\author{Thibaut Gallet}

\author{Conrad Spindler}

\author{Omar Ramírez}

\author{Christian Kameni Boumenou}


\author{Susanne Siebentritt}

\author{Alex Redinger}
\ead{alex.redinger@uni.lu}

\address{Department of Physics and Materials Science, University of Luxembourg, Luxembourg}

    \begin{abstract}
An in-depth understanding of the electronic properties of grain boundaries (GB) in polycrystalline semiconductor absorbers is of high importance since their charge carrier recombination rates may be very high and hence limit the solar cell device performance. 
Kelvin Probe Force Microscopy (KPFM) is the method of choice to investigate GB band bending on the nanometer scale and thereby helps to develop passivation strategies. 
Here, it is shown that amplitude modulation AM-KPFM, which is by far the most common KPFM measurement mode, is not suitable to measure workfunction variations at GBs on rough samples, such as Cu(In,Ga)Se$_2$ and CH$_3$NH$_3$PbI$_3$. This is a direct consequence of a change in the cantilever-sample distance that varies on rough samples.

Furthermore, we critically discuss the impact of different environments (air versus vacuum) and show that air exposure alters the GB and facet contrast, which leads to erroneous interpretations of the GB physics. 

Frequency modulation FM-KPFM measurements on non-air-exposed CIGSe and perovskite absorbers show that the amount of band bending measured at the GB is negligible and that the electronic landscape of the semiconductor surface is dominated by facet-related contrast due to the polycrystalline nature of the absorbers. 

\end{abstract}


\end{frontmatter}



    \section{Introduction}
    \label{S:1}
High performance polycrystalline thin film solar cells are very interesting alternatives to the mainstream silicon-based devices due to their low manufacturing costs combined with low energy payback time \cite{Powalla2017AdvancesCuInGaSe2}. 
Stacks of different materials, that form the solar cell are usually deposited on low cost glass substrates, flexible polyamide or steel foils, which leads to polycrystalline absorbers with typical grain sizes in the micrometer range, separated by grain boundaries (GBs). The optoelectronic impact of these planar lattice defects in thin-film solar cells is still intensively studied \cite{Krause2020,Hieulle2018ScanningCells}

It is known that the interruption of the crystal periodicity in semiconductor materials and the resulting dangling or strained bonds often lead to energy states in the bandgap, which thereby may lead to the formation of charges at the GBs \cite{Rau2009GrainCells}. According to simulations, charges at the GBs deteriorate the solar cell performance due to losses in open-circuit voltage ($V_\mathrm{OC}$) and fill factor ($FF$) \cite{Rau2009GrainCells, taretto_numerical_2008}. Consequently, from a simulation point of view, GBs are not beneficial as they act as recombination centers for charge carriers \cite{Saniz2017StructuralCuInSe2}. 

Measurement techniques with a high spatial resolution, combined with the ability to measure the electronic landscape are indispensable to understand the GB physics.
Examples of techniques that are commonly used in this field are: scanning and transmission electron microscopy (SEM/TEM) \cite{abou-ras_direct_2012,Abou-Ras2012ConfinedSemiconductors,Abou-Ras2019NoCells,Zhou2019}, atom probe tomography \cite{Rivas2020,cojocaru-miredin_characterization_2011, Schwarz2020OnFilms, Keller2013GrainTomography}, photoluminescence techniques \cite{Spindler2019ElectronicModel,DeQuilettes2016}, and scanning probe techniques \cite{Hieulle2018ScanningCells}, such as scanning near-field optical microscopy \cite{Neumann2015PhotoluminescenceLimit}, scanning tunneling microscopy \cite{Boumenou2020PassivationTreatment, Broker2015CorrelatingSurfaces} and Kelvin probe force microscopy (KPFM) \cite{Gallet2021Co-evaporationStability, Hanna2006TextureFilms, Sadewasser2002High-resolutionSurfaces,Leblebici2016}.

In this work, we focus on one particular scanning probe technique, namely Kelvin Probe Force Microscopy  \cite{nonnenmacher_kelvin_1991, Jacobs1997SurfaceSPM}, which has been used frequently to measure the electrostatic properties of thin film solar cell absorbers \cite{Melitz2011KelvinApplication, Hidalgo2019ImagingCells, Hieulle2018ScanningCells}. During KPFM, the contact potential difference (CPD) between a sharp conductive probe and the surface of a sample can be acquired with nanometer resolution. The measured CPD is proportional to the workfunction difference between the probe and the sample. In the case of a known probe workfunction, KPFM allows to quantify band bending at grain boundaries in polycrystalline materials \cite{Melitz2010ScanningSurfaces, Sadewasser2002High-resolutionSurfaces, Baier2012TowardDistributions, Leendertz2006EvaluationFilms, Jiang2004LocalFilms}. However, as we will show in this manuscript, the correct interpretation of the measured workfunction changes at the GBs is not trivial and many pitfalls need to be considered and eliminated. 

In this work, we will show explicitly how different measurement setups and sample properties impact the interpretation of band bending at GBs. We focus on two very important classes of materials, namely halide perovskites and Cu(In,Ga)Se$_2$ (CIGSe).
Both material systems have demonstrated power conversion efficiencies (PCE) higher than 23\% over the last years \cite{Green2021Solar57} and are already industrially manufactured or are in the process of entering the market. 
Due to the abundance of GBs in these materials, recombination losses are often attributed to recombination at the GBs. Quite a few results in literature suggest a strong relation between charge accumulation and variations in power conversion efficiency (PCE). Specifically there, KPFM emerged as a powerful tool to access the local electronic properties at the sample surface  \cite{Melitz2011KelvinApplication, Sadewasser2002High-resolutionSurfaces, Jiang2004LocalFilms, nicoara_effect_2017, Jiang2004DoesDevice, Hafemeister2010LargeFilms, Shao2016GrainFilms, Yun2015BenefitCells, Siebentritt2006EvidenceChalcopyrites}. In the following, we compare the results for two KPFM detection methods: {\bf A}mplitude {\bf M}odulation KPFM ({\bf AM}-KPFM) under ambient conditions and {\bf F}requency {\bf M}odulation KPFM ({\bf FM}-KPFM) under ultra-high vacuum  (UHV) conditions. Even though the AM-KPFM is known to suffer from several issues \cite{polak_note_2014, charrier_real_2008, Axt2018KnowDevices}, it is still the most widely used method in the solar cell community.  We  compare the two measurement modes on four types of samples: gold on silicon acting as a reference sample, single-crystalline Cu-rich CISe, polycrystalline Cu-poor CIGSe and polycrystalline CH$_3$NH$_3$PbI$_3$ (MAPI). Our measurements are supplemented with electrostatic calculations to reveal how the surface roughness impacts the KPFM results. Finally, we discuss grain boundary band bending measurements that are free of artifacts.

    \section{Materials and Methods}
    \label{S:2}

    \subsection{Sample preparation}

\textbf{\textit{Perovskite \& CIGSe absorbers}}: Methylammonium lead triiodide perovskites (CH$_3$NH$_3$PbI$_3$) (MAPI) absorbers were deposited on FTO (fluorine-doped tin oxide) covered glass substrates  via co-evaporation carried out in a physical vapour deposition chamber embedded in a nitrogen-filled glovebox. A constant temperature of 330\,$^{\circ}$C was used to evaporate PbI$_2$ and the temperature of the MAI was kept at 110\,$^{\circ}$C. The growth was carried out at room temperature. The samples were then transferred to the UHV KPFM apparatus, using an inert-gas transfer system.

Epitaxial Cu-rich CuInSe$_2$ (CISe) films were grown by metal-organic vapor phase epitaxy (MOVPE) on (100)-oriented semi-insulating GaAs wafers at 530\,$^{\circ}$C and 50\,mbar. Details of the process can be found in \cite{Spindler2019ElectronicModel,Ramirez2021}. The absorbers were approximately 500\,nm thick and the samples were transferred with the same suitcase directly into the SPM chamber without air exposure. A Cu/In=1.15 was measured via energy dispersive X-Ray analysis (EDX) at 10\,kV.

Polycrystalline CIGSe absorbers were grown via multi-stage coevaporation carried out in a molecular beam epitaxy system (MBE). Details of the growth process can be found in \cite{Babbe2016QuasiLayers}. The Cu/(In+Ga) and Ga/(Ga+In) ratios measured by EDX were 0.85 and 0.31 respectively. The quasi-Fermi level splitting measured via calibrated photoluminescence was 697\,meV under one sun equivalent conditions, which corroborates that the absorbers had an excellent optoelectronic quality. The samples were transferred from the MBE chamber (base pressure low $10^{-9}$\,mbar range) to the SPM chamber via a UHV suitcase. 

    \subsection{AFM characterization techniques}

FM-KPFM measurements were carried out using a UHV VT-AFM system (Omicron) operated in the low 10$^{-11}$\,mbar pressure range. 
Topography and potential information were measured simultaneously by using the cantilever resonance frequency as feedback with two independent lock-in amplifiers. 
The applied AC voltage ranged from 0.2\,V to 0.4\,V at 1.25\,kHz. The used probes were Pt/Ir PPP-EFM (nanosensors) with a resonance frequency between 70\,kHz and 90\,kHz. 

AM-KPFM measurements were carried out with a  Nanoscope V (Digital instruments) operated in double-pass mode, with topography being acquired in the first pass at the resonance frequency of the probe. Surface potential was acquired in the second pass with the probe lifted by a few nanometers from the sample surface. 
The AC voltage used during the KPFM measurements was 5\,V. The used probes were identical to the ones for the UHV FM-KPFM measurements. 

The differences between AM- and FM-KPFM are discussed in more detail in the Supplementary Information (SI). The experimental implementation and their limitations are explained and illustrated with measurements on reference samples (see \textbf{Figs. \ref{FS:1}} and \textbf{\ref{FS:2}} of SI). 
From the measurements carried out on the reference sample one can conclude that AM-KPFM is easier to implement and that this method has a better voltage resolution, down to 5\,mV, while FM-KPFM can detect CPD values down to 10\,mV \cite{Melitz2011KelvinApplication}. On the other hand, FM-KPFM yields to a more precise quantification of the workfunction and better lateral resolution in complete agreement with the literature (see detailed discussion in SI). 


\section{Literature Review on band banding and solar cell efficiencies}

\begin{figure}[ht]
    \centering
    \includegraphics[width=0.85\linewidth]{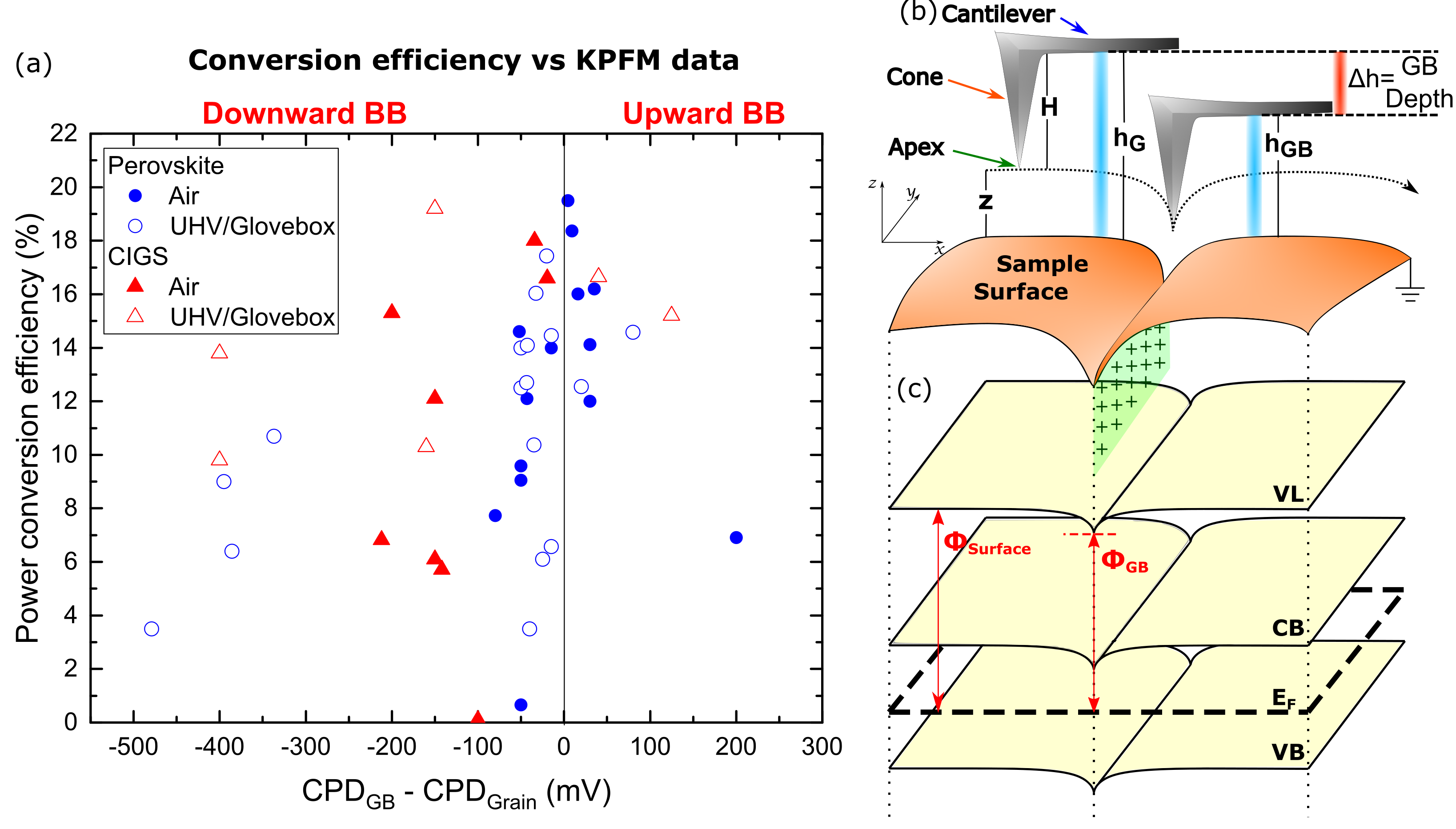}
    \caption{(a) Contact potential differences $CPD_\mathrm{GB}$ – $CPD_\mathrm{Grain}$  at the GBs and at the grains for MAPI and CIGSe absorbers as a function of the reported PCE. (b) Schematic representation of the scanning mechanism during KPFM measurements on GBs, illustrating that for a constant probe-sample distance $z$, the cantilever-sample distance changes when entering the GB. 
    Blue, red and green arrows depict the cantilever, cone and probe apex of the SPM probe with $H$ defining the cone height.
    The distances between the cantilever and the sample are marked with blue lines, $h_{G}$ and $h_{GB}$ represent the distance when the probe apex is at the grain and at the GB, respectively.
    The GB depth is defined as $\Delta h = h_{G} - h_{GB}$.
    (c) 3D illustration of the band diagram showing the vacuum level ($VL$), conduction band ($CB$), Fermi level ($E_\mathrm{F}$) and valence band ($VB$).}
    \label{F:1}
\end{figure}

Before presenting the results, we review the available literature on band bending at grain boundaries measured with KPFM. We limit the discussion to MAPI and CIGSe absorbers and their devices. 
As discussed in the introduction and in the SI, band bending can be characterized by the difference in workfunction (or CPD) between the GBs and the grains. 
In order to see if there is a direct correlation between band bending and solar cell efficiency, we plot in \textbf{Fig.~\ref{F:1}\,(a)},  the difference between the reported CPD values at GBs ($CPD_\mathrm{GB}$) and the CPD at the grain surfaces ($CPD_\mathrm{Grain}$) as a function of the reported PCE for devices made from the same absorber \cite{Jiang2004LocalFilms, nicoara_effect_2017, Yun2015BenefitCells, dymshits_electronic_2015, ma_temperature-dependent_2019, chen_controllable_2014, chen_optoelectronic_2015, kim_effects_2016, edri_why_2014, chang_enhancing_2016, jiang_enhancing_2014, kim_efficient_2015, li_influence_2016, adhikari_interfacial_2015, Kim2019TheGrading, kim_simple_2019-1, Hanna2006TextureFilms, Nicoara2019DirectTreatments, adhikari_nanoscale_2016,Bahrami2020NanoscaleCells, Hoque2018EffectsCells,Li2020SurfaceProperties}. 
All displayed values are plotted such that they refer to measurements where the bias is applied to the sample. In the SI (\textbf{Table \ref{TS:1}}) more experimental details for each of the references can be found. Negative values for $CPD_\mathrm{GB}$ – $CPD_\mathrm{Grain}$ refer to downward band bending due to an excess of positive charges at the GBs whereas positive values for $CPD_\mathrm{GB}$ – $CPD_\mathrm{Grain}$ refer to upward band bending accompanied with an excess of negative charges. This is illustrated schematically in \textbf{Fig.~\ref{F:1}\,(b,c)} where a cantilever and a representation of a surface including a positively charged GB is presented together with the corresponding band diagram. Charged defects that are responsible for band bending impact the vacuum level (VL), the conduction band (CB), and the valence band (VB) compared to the grain interior.

In \textbf{Fig.~\ref{F:1}\,(a)} the data points are split into different categories namely  air-exposed samples (solid symbols) and samples measured under UHV or inert gas conditions (open symbols). 
From the graph,  a preference for downward band bending can be identified for both material systems. 
Additionally, most of the CPD data ranges from -150\,mV to 150\,mV and the inert gas and UHV measurements suggest stronger band bending.
However, we do not see an obvious correlation between PV performance and band bending. The question, if a certain GB band bending is favoured for high efficiency devices can therefore not be answered from this graph. 

From a device point of view this result is expected since losses in solar cell performance can have multiple reasons, not directly linked to GBs. Examples include, but are not limited to, unfavourable doping levels and band alignments, shunts, and high non-radiative recombination rates (see for example: \citet{Scheer2011ChalcogenideDevices}). 

However, there are other reasons that are not directly linked to the absorber and the fabrication process itself, which will be discussed in the following. First of all, many KPFM measurements presented in \textbf{Fig.~\ref{F:1}\,(a)}  were performed under ambient conditions, which might induce “environmental” artifacts such as sample contamination with oxygen and water, modifying  the surface dipoles \cite{Sugimura2002PotentialMicroscopy, Salomao2015DeterminationMicroscopy, burgo_electric_2011}. 
Secondly, most of the KPFM data shown in \textbf{Fig.~\ref{F:1}\,(a)} (check \textbf{table \ref{TS:1}}) were collected using AM-KPFM, which is the KPFM variant known in literature to be strongly influenced by the long range electrostatic force contribution of the complete probe to the measured electric force \cite{Colchero2001ResolutionMicroscopy, Wagner2015KelvinDevices, Axt2018KnowDevices,Glatzel2003AmplitudeMicroscopy, Zerweck2005AccuracyMicroscopy, charrier_real_2008, Melitz2011KelvinApplication, Diesinger2012CapacitiveKPFM}. 
The direct consequence of this coupling is  a poor spatial resolution and a wrong absolute workfunction value (see SI for measurements performed on a reference sample). This problem has been known in the community for quite some time. Here we would like to highlight an additional problem when measuring polycrystalline solar cell absorbers with AM-KPFM, namely the surface roughness. 

In \textbf{Fig.~\ref{F:1}\,(b)} a schematic representation of a probe moving over a  GB is depicted. 
When the probe scans over the surface, the sharp probe apex follows the contour of the topography, thereby also entering into the GB region. 
As a consequence, the distance between the cantilever and the sample surface changes by $\Delta h = h_{G} - h_{GB}$, where $h_{G}$ and $h_{GB}$ denote the cantilever-sample distance at the top of the grain and at the GB, respectively. 
This change has so far never been taken into account when discussing KPFM measurements on polycrystalline absorbers, since typical values for the probe height $H$ are in the order of 12\,$\mu$m and $\Delta h$ values only range up to several hundred nanometers. 
Consequently, it seems at first glance well justified that these changes can be neglected. 
As we will show in the following, this assumption is not justified for the case of AM-KPFM whereas it works out well for the FM-KPFM case, which has important consequences for the interpretation of work function changes at the GBs.

In the rest of this manuscript, we therefore investigate the impact of the environment and the impact of the grain boundary depth $\Delta h$ on GB band bending in order to better understand the literature data and our own measurements.

\section{The impact of the environment and sample history}
    \label{PeroEnvir}
  
\begin{figure}[H]
    \centering\includegraphics[width=0.8\linewidth]{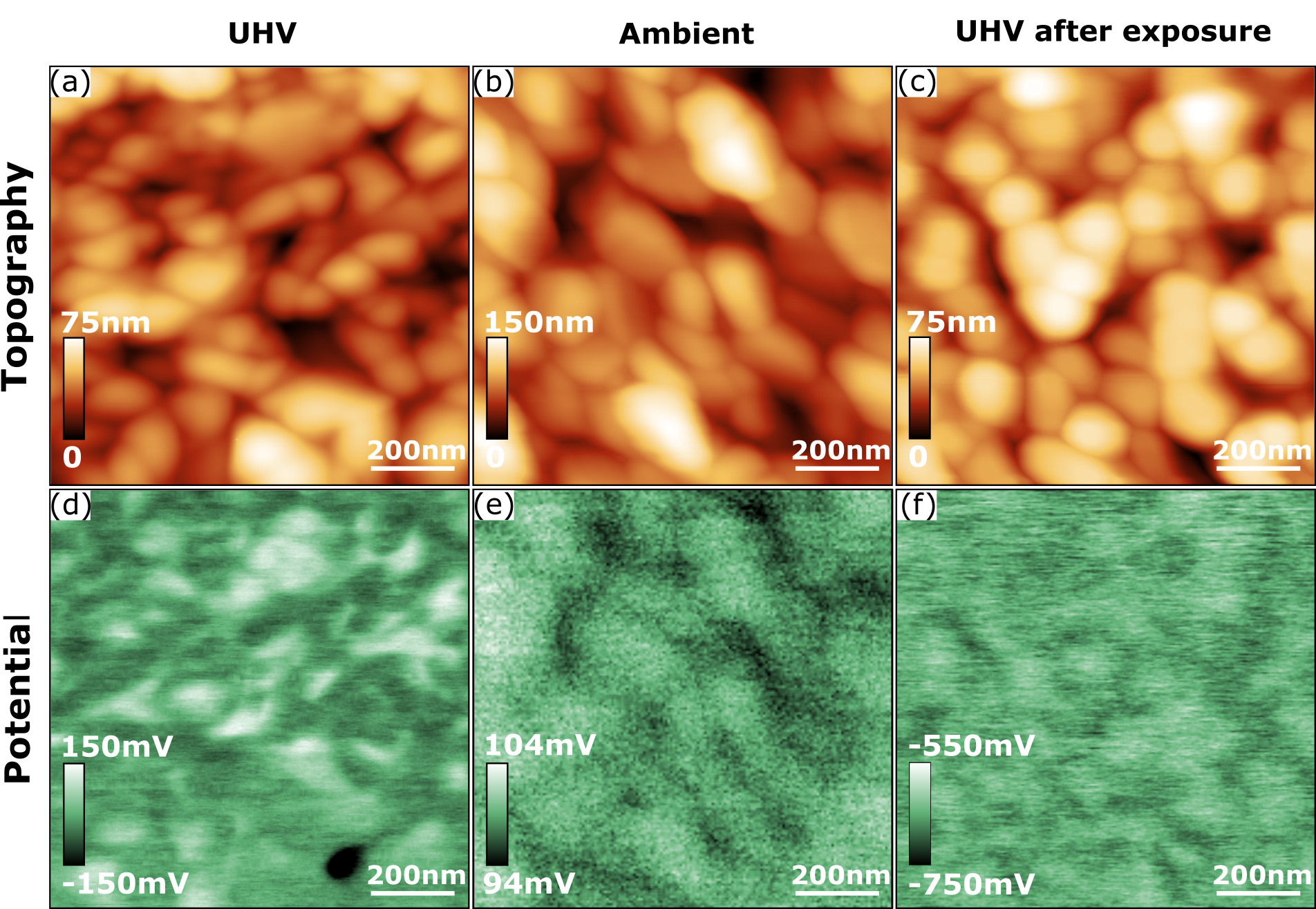}
    \caption{ KPFM data for a  MAPI showing the topography (a-c) with the respective CPD maps (d-f) measured under UHV (FM-KPFM), ambient (AM-KPFM) and UHV (FM-KPFM) after exposing it to air. 
    Topography and CPD maps are shown with an optimum contrast illustrated with a scale bar in the respective image.
    Facet-related contrast completely vanished after exposing the sample to ambient conditions.}
    \label{F:2}
\end{figure}

In \textbf{Fig.~\ref{F:2}} FM- and AM-KPFM  results carried out under UHV and under ambient conditions on the \underline{same} MAPI absorber are presented. FM-KPFM measurements were first performed under UHV on a non-air-exposed  perovskite sample (\textbf{Fig.~\ref{F:2}\,(a,d)}), then transferred to the air AM-KPFM setup to measure it under ambient conditions (\textbf{Fig.~\ref{F:2}\,(b,e))}. 
After two days of air exposure the sample was introduced back into the UHV system in order to check if the same topography and workfunction map could be recovered (\textbf{Fig.~\ref{F:2}\,(c,f)}).
All the images were adjusted for maximum contrast and are reprinted in the SI with the same scale bar contrast (\textbf{Fig.~\ref{FS:4}}).

FM-KPFM measurements under UHV conditions (\textbf{Fig.~\ref{F:2}\,(a,d)}) clearly displayed facet-dependent features in the workfunction map, where variations of around 250\,mV between facets could be measured. 
\textbf{Fig.~\ref{FS:6}} in the SI shows topography and workfunction line profiles that corroborate the facet-related contrast. 
Comparing the measurements in UHV with FM-KPFM to the AM-KPFM in air we observed one important difference. The surface potential contrast almost completely vanished, as indicated by the scale bar value of \textbf{Fig.~\ref{F:2}\,(e)}. 
Only a surface potential contrast in the order of 10\,mV was observed, which was almost exclusively located at the grain boundaries.
We note that this value, although small, is in agreement with \textbf{Fig.~\ref{F:1}\,(a)} where most of the measurements performed with an AM-KPFM setup in air showed only small variations at the GBs.
Interestingly, we noticed that the deeper the GB, the higher the KPFM contrast was, as shown in the \textbf{Fig.~\ref{FS:9}} of SI.  

To check the reversibility of the observed changes, the sample was moved back to UHV to perform once again FM-KPFM measurements on the same, albeit air-exposed sample (\textbf{Fig.~\ref{F:2}\,(c, f)}). 
The magnitude of the CPD values increased again, but the CPD contrast prior to air exposure could not be recovered. 
This time, a weak correlation of the CPD contrast at grain boundaries and no facet-dependent contrast was observed. 

The results presented in \textbf{Fig.~\ref{F:2}\,(a,d}) and \textbf{Fig.~\ref{F:2}\,(c, f)} showed that the workfunction maps measured with FM-KPFM in UHV depend sensitively on the sample history. 
The results can be attributed to a partial or total oxidation/decomposition of the perovskite surface, which generated nonuniform and random patterns in the workfunction map. 
\textbf{Figs. \ref{FS:7} and \ref{FS:8}} of SI show how the topography and the workfunction of the MAPI sample change over time when exposed to air. 
Over time, a continuous increase of CPD and changes in morphology are observed when measurements are carried out under ambient conditions whereas almost stable CPD and topography is observed under UHV.
It is therefore indispensable that KPFM measurements are carried out on samples without air exposure. 

Therefore, the AM-KPFM results measured in air are different than the FM-KPFM measurements in UHV. The average workfunction and even more important the GB contrast observed in both methods differ substantially. Consequently, the band bending information extracted from the two measurements are different, which renders a direct correlation with the solar cell performance challenging. However, as we will show in the following, there is a much more fundamental difference between the two measurement modes, which results in an erroneous GB contrast in AM-KPFM.  

    \subsection{The impact of KPFM mode on Perovskite GB contrast}
    \label{PeroKPFMmode}

\begin{figure}[H]
    \centering\includegraphics[width=0.8\linewidth]{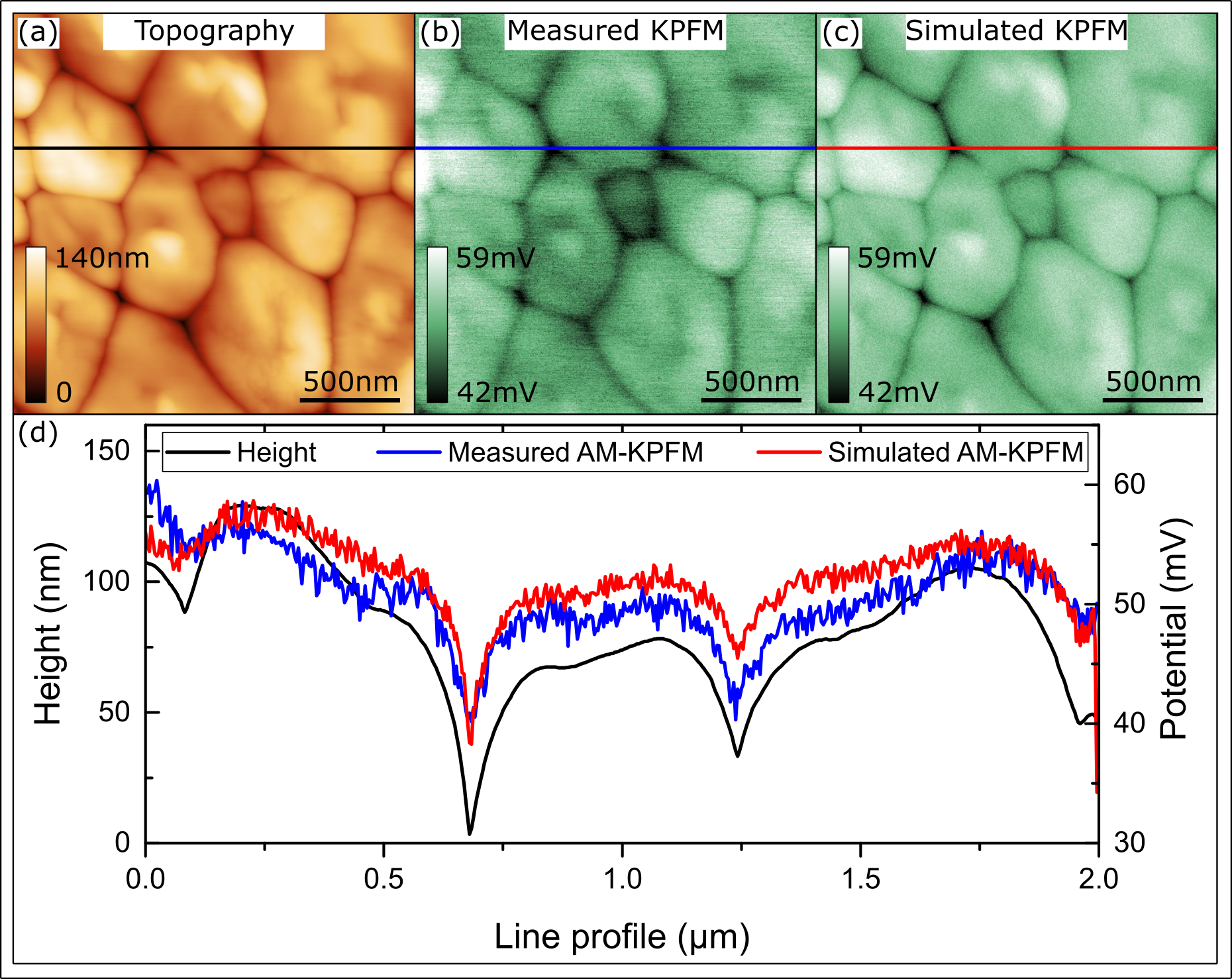}
    \caption{ (a) topography and (b) workfunction map of a MAPI absorber measured with AM-KPFM under ambient conditions.
    (c) calculated AM-KPFM image taking into account only the changes in the electrostatic force caused by the surface roughness (details see text). 
    (d) line profiles from the topography, from the measured and calculated AM-KPFM image.}
    \label{F:3}
\end{figure}

In the following, we discuss why the measured contrast at the grain boundaries acquired with AM-KPFM cannot be trusted. This is shown exemplary in \textbf{Fig.~\ref{F:3}}, where the topography and CPD images measured on another MAPI absorber are depicted. Again, a clear correlation between grain and grain boundaries is observed,  which could be attributed to a small downward band bending. We stress that all other AM-KPFM measurements on other perovskite absorbers could also have been used for the following discussion (see \textbf{Fig.~\ref{FS:5}} where the following evaluation  is done on the measurement presented in \textbf{Fig.~\ref{F:2}}).   

The total force acting between the cantilever and the sample can be divided into several contributions, namely, the cantilever, the cone, and the probe apex, as illustrated in the sketch of \textbf{Fig.~\ref{F:1}\,(b)}. 
It has been shown for probe-sample distances that are relevant in most experiments that the cantilever contribution in AM-KPFM dominates \cite{Wagner2015KelvinDevices}.  
In \textbf{Fig.~\ref{FS:3}}, the individual contributions (apex, cone, cantilever) are summarized, as it was done previously by \citet{Wagner2015KelvinDevices}.
All measurements presented in this manuscript were carried out with lift-heights of 50\,nm. In that case, the dominant contribution to the electrostatic force is given by the cantilever  (denoted $F_{\mathrm{lever}}$), which can be expressed via: 
\begin{equation}
    F_{lever}=-\frac{1}{2}\frac{\epsilon_0 A}{(z+H)^2}U_{ts}^2=-\frac{1}{2}\frac{\partial C}{\partial r} U_{ts}^2
\label{eq1}
\end{equation}

The cone height is denoted as $H$ and the distance between the probe apex and the sample as $z$. The potential difference between the cantilever and the sample is given as $U_{\mathrm{ts}}$, $A$ is the area of the cantilever and $\epsilon_0$ the permittivity. 
Consequently, the force exhibits a quadratic dependence on $U_{\mathrm{ts}}$ and the remaining distance dependence can be lumped in the capacitance gradient $\frac{\partial C}{\partial r} $, which is assumed to be constant during AM-KPFM.
As sketched in \textbf{Fig.~\ref{F:1}\,(b)} the measurements of GBs at a constant probe-sample distance $z$ lead to a change in the distance between the cantilever and the sample, reducing from $h_G$ when the probe apex is at the grain to $h_{GB}$ when the probe apex enters the GB. 
We denoted this change as $\Delta h = h_{G} - h_{GB}$, which results in a change in the electrostatic force $\Delta F_{\mathrm{lever}}$, which can be calculated via:
\begin{equation}
    \Delta F_{lever}=-\frac{1}{2}\frac{\epsilon_0 A}{(z+H-\Delta h)^2}U_{ts}^2+\frac{1}{2}\frac{\epsilon_0 A}{(z+H)^2}U_{ts}^2\equiv-\frac{1}{2}\frac{\epsilon_0 A}{(z+H)^2}\Delta U_{ts}^2
    \label{eq2}
\end{equation}

Equation \ref{eq2} is written in two different forms. In the first part, a change in force is reflected in a change of the capacitance gradient, which is just the difference between the probe apex inside the GBs and the probe apex on the grain surface. 
The expression after the equivalent sign arises from the fact that in AM-KPFM  the capacitance gradient is usually not compensated and assumed to be constant over the scanned area. 
Consequently, changes in force due to changes in height are not explicitly taken into account. 
Therefore, all changes in $\Delta F_{\mathrm{lever}}$ in AM-KPFM result in an apparent  potential variation, denoted as $\Delta U_{ts}$ with $\frac{\partial C}{\partial r} $ being constant  (see equation \ref{eq2}). Solving for $\Delta U_{ts}$ therefore allows to estimate the apparent change in CPD, induced by a change of the cantilever-sample distance. 
In other words, $\Delta U_{ts}$ shows the variation in the electrostatic force due to changes in the capacitive coupling when the distance between the cantilever and the sample is changed, keeping the probe apex/sample distance constant.
\begin{equation}
    \Delta U_{ts}=U_{ts} \sqrt{\frac{2(z+H)\Delta h -(\Delta h)^2}{(z+H-\Delta h)^2}}
    \label{eq3}
\end{equation}

The experimentally measured averaged CPD value in \textbf{Fig.~\ref{F:3}\,(b)} was $U_{ts} = 0.05$\,V, the lift height during KPFM was set to $z=50\times10^{-9}$\,m, the cone height $H=12\times10^{-6}$\,m, and the average GB depth $\Delta h=100\times10^{-9}$\,m. 
This leads to an apparent change in the CPD by $\Delta U_{ts}=6.4$\,mV. This value is very close to the contrast variations we observed in \textbf{Fig.~\ref{F:3}\,(b)}.

However, each GB has a slightly different depth. To take this into account, we used the topography acquired during the AM-KPFM (\textbf{Fig.~\ref{F:3}\,(a)}) as the input value for changes in the cantilever-sample distance $\Delta h$. 
Then for each pixel of the topography image, we calculated $\Delta U_{ts}$, which is depicted in \textbf{Fig.~\ref{F:3}\,(c)}. 
Random noise of 2\,mV peak to peak was added to this image in order to be as close as possible to the real KPFM setup. \textbf{Fig.~\ref{FS:10}} of SI shows a comparison between the simulated images with and without the random noise.

The results of the calculation highlight impressively that the measured workfunction map via AM-KPFM is almost identical to the simulated image that considers only changes in electrostatic forces due to variations in the cantilever-sample distance at the GBs. 
There is no information included in the calculation that arises from the probe apex or cone. Consequently, the calculation suggests that there is no physical meaning in the AM-KPFM measurement performed on the perovskite sample. 
The only remaining explanation that would allow us to interpret the AM-KPFM measurements as band bending would be to assume that the number of defects at the GBs would be proportional to the GB depth. 
To exclude this, measurements on a rough single-crystalline surface without GBs need to be analyzed, which is discussed in the next section.

    \subsection{Facet contrast measured on single-crystalline CISe}
    \label{CISe}

\begin{figure}[H]
    \centering\includegraphics[width=0.8\linewidth]{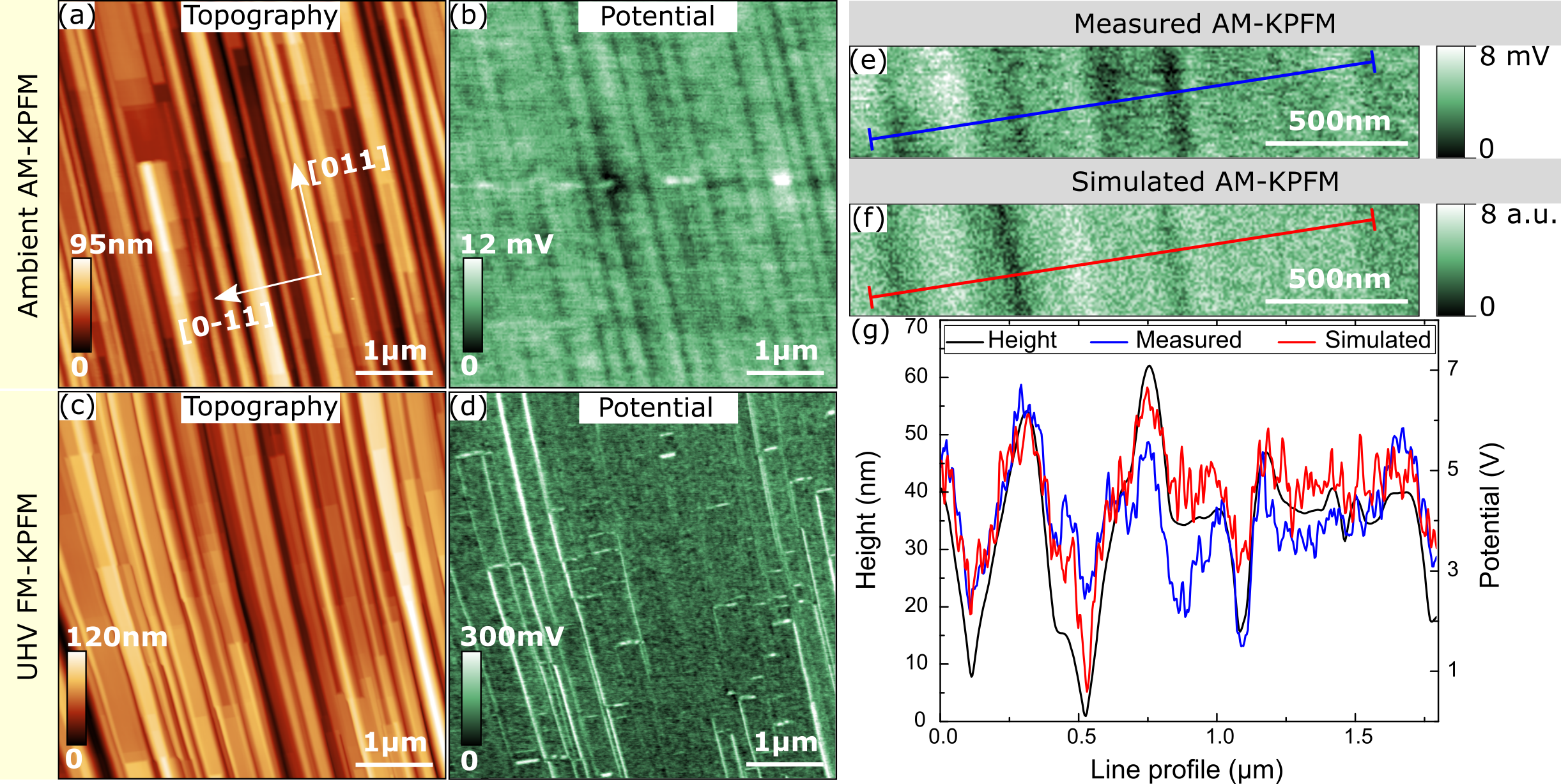}
    \caption{Ambient AM-KPFM, UHV FM-KPFM and simulated AM-KPFM acquired on an epitaxially grown CISe absorber. 
    (a,b) AM-KPFM under ambient conditions, where (a) shows the topography and (b) shows the surface potential images. 
    (c,d) FM-KPFM under UHV conditions, where (c) shows the topography and (d) shows the surface potential. 
    (e) measured and (f) calculated AM-KPFM (details see text). (g) line profiles of the topography, of the measured and calculated AM-KPFM measurement.}
    \label{F:4}
\end{figure}

In analogy to the measurements carried out on the MAPI absorbers, ambient AM-KPFM and UHV FM-KPFM  were acquired on the \underline{same} epitaxial CISe surface.
\textbf{Fig.~\ref{F:4}\,(a,c)} depict the topography of the layers, where trenches aligned in the [011] direction with height variations in the order of 100\,nm between valley and peak were observed. 
This highly textured growth is well known for epitaxially grown CISe material and it is related to the formation of polar (112) facets \cite{Jaffe2003Defect-induced2, Siebentritt2006StabilitySystem}. 
\textbf{Fig.~\ref{F:4}\,(b,d)} depicts the respective surface potential images adjusted for maximum contrast (for the same images with fixed scale bars from 0 to 300\,mV, check SI \textbf{Fig.~\ref{FS:11} (b-d)}).
Similar to the AM- versus FM-KPFM measurements on MAPI, a large difference in the CPD variations was observed for both measurement techniques, with valley to peak values of only 12\,mV in AM-KPFM and up to 300\,mV in FM-KPFM.

In the FM-case (\textbf{Fig.~\ref{F:4}\,(d)}) some topographic features, namely, the (112) and (312) facets showed higher workfunction values compared to the other facets as reported previously \cite{lanzoni_surface_2019}.
In AM-KPFM (\textbf{Fig.~\ref{F:4}\,(b)}) almost all topographic features exhibit a weak contrast in the potential images, which is not in accordance with the FM-case. 
To understand the origin of the measured AM-KPFM signal, we used the same approach, as described in the \textbf{section \ref{PeroKPFMmode}} and calculated the changes in CPD arising only from the change in height between the cantilever and the sample. 
\textbf{Fig.~\ref{F:4}\,(e,f)} depict the measured and the calculated AM-KPFM images from the same area of the single-crystal. 
Both images are very similar, with the bright and the dark regions at the same lateral positions. Line profiles with 20 pixels thickness from the measured (blue), simulated (red), and topography (black) images were taken from the same spot and are presented in \textbf{Fig.~\ref{F:4}\,(g)}. 
The agreement between the simulated and the measured data is excellent, which allows to conclude that the signal in AM-KPFM arose almost entirely from the capacitive coupling between the cantilever and sample. 

The measurements on epitaxial CISe and measurements on polycrystralline MAPI therefore show strong capacitive cross-talk at the trenches. While for polycrystalline material, discussed in \textbf{section \ref{PeroKPFMmode}}, a potential mixing of cross-talk and downward band bending could not be excluded, the measurements on epitaxial CISe, where GBs were absent unambiguously show that the height changes in AM-KPFM lead to a strong cross-talk. 
Consequently, AM-KPFM measurements in double pass mode cannot be used to measure the band bending at GBs. Intuitively the same calculations could be repeated for FM-KPFM. However, as shown in \textbf{Fig.~\ref{FS:3}}, the cantilever contribution for all tip samples distances used experimentally is negligible. Only the tip apex and the cone contribute significantly to the electrostatic force gradient used to measure the CPD in FM-KPFM. Consequently, the use of \textbf{equation \ref{eq3}} to calculate a FM-KPFM image is not meaningful. The cone and apex contributions do not depend on $\Delta h$ and will therefore not be influenced by the surface roughness (see equations in SI for details).


    \subsection{Surface contamination and polycrystalline absorbers} 

So far, we have identified two important problems that need to be taken care of to measure accurate workfunction maps. 
Samples cannot be air-exposed for a prolonged time and FM-KPFM measurements instead of AM-KPFM measurements need to be performed in order to reduce the capacitive cross-talk at the GBs. 
In this section, the aspect of surface contamination will be discussed in more detail and measurements on non-air-exposed polycrystalline CIGSe absorbers will be analyzed. 

\begin{figure}[H]
    \centering\includegraphics[width=0.8\linewidth]{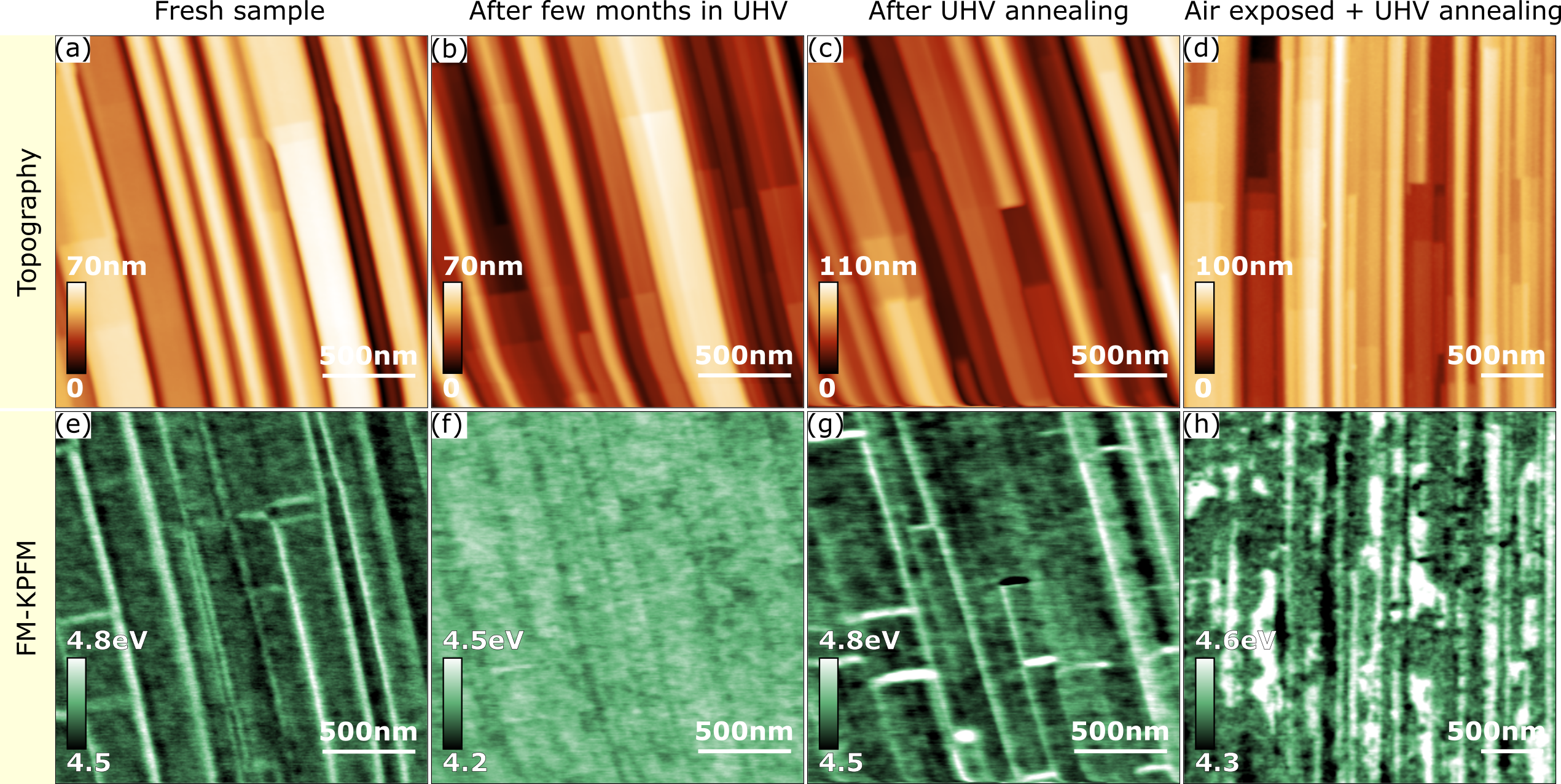}
    \caption{FM-KPFM results on an epitaxial CISe sample exposed to different levels of surface contamination.
    (a-d) topography and (e-h) workfunction maps.
    (a) and (e) represent the as-grown and non-air-exposed condition.
    (b), (f) represent the same non-air-exposed sample that was kept in the UHV environment for 20\,months.
    (c) and (g) show topography and KPFM measurements after UHV annealing of the aged sample ((b),(f)).
    (d) and (h) depict the topography and workfunction map after UHV annealing of another piece of the same sample that was exposed to ambient conditions for 6 months and then annealed.}
    \label{F:5}
\end{figure}

\textbf{Fig.~\ref{F:5}} shows topography (a-d) and workfunction maps (e-h) of the \underline{same} single-crystal CISe absorber measured with different levels of surface contamination. 
The  fresh sample exhibited the facet-dependent contrast, as discussed in the \textbf{section \ref{CISe}} and detailed in \citet{lanzoni_surface_2019}. 
After 20 months in UHV, this sample was measured again (\textbf{Fig.~\ref{F:5}\,(b,f)}) and despite the fact that the topography did nominally not change, the high workfunction values at some specific facets were no longer present.
Instead, only a shallow dark workfunction contrast of approximately 30\,meV aligned in the direction of the trenches was visible. 
The origin of this remaining contrast is not clear at present. The physical processes that govern the disappearance of the facet contrast is most likely linked to adsorbates that physisorbed on the absorber surface and obscured the surface dipole induced workfunction contrast \cite{Kahn2016FermiLevel}. In order to clean the substrate surface, the sample was annealed for 30\,minutes at 200\,$^\circ$C in UHV. 
The annealing temperature was chosen such that absorber surface deterioration due to loss of selenium could be neglected  \cite{Boumenou2020PassivationTreatment, Monig2014HeatCells}.

\textbf{Fig.~\ref{F:5}\,(c-g)} depict the FM-KPFM measurement after the annealing procedure. 
Again, the topography image did not change compared to the previous measurements (\textbf{Fig.~\ref{F:5}\,(a,b)}), however, the surface potential map showed again the bright facet contrast that was very characteristic for the fresh sample. 
The same trend was also observed for the average workfunction values, where a recovery to the initial value was observed (fresh sample: $4.62\pm 0.04\,eV$, adsorbate covered surface: $4.36\pm0.03\,eV$, UHV annealed sample: $4.60\pm0.06\,eV$).
These results corroborated that adsorbates-induced changes of the CISe surface did not lead to a permanent modification of the electrostatic landscape of the semiconductor and a recovery of the initial CISe surface properties was possible via a mild UHV annealing.  

We did the same annealing procedure for another piece of the \underline{same} sample that was stored in air instead of UHV. In \textbf{Fig.~\ref{F:5}\,(d,h)}, the topography and workfunction map of an air-exposed CISe sample after heating are presented. 
A clear contrast in CPD was observed, which is however not linked to a specific facet but rather to a random distribution of areas with varying workfunction values. 
We associated it to the formation of an oxide layer on the CIGS surface that cannot be removed with the annealing for 30 minutes at 200\,$^\circ$C in UHV. 
The average workfunction value of this sample ($4.42\pm 0.08\,eV$) was lower than the one of the pristine absorber, which corroborated that annealing of air-exposed samples could not be used to recover the initial workfunction values and the facet related contrast. 

Unfortunately, the same procedure cannot be used for the case of halide perovskites, which are very temperature sensitive. 
In a recent work from \citet{Gallet2021Co-evaporationStability}, this was corroborated for the case of MAPI absorbers where in the best case, the samples could be heated up to approximately 100\,$^\circ$C in nitrogen. In vacuum, this value decreases even further, due to the volatile nature of the MAI \cite{Ro2020Co-EvaporatedLayer}. Consequently, for the case of halide perovskites a fast measurement directly after growth without air exposure is mandatory. 

\begin{figure}[H]
    \centering\includegraphics[width=0.8\linewidth]{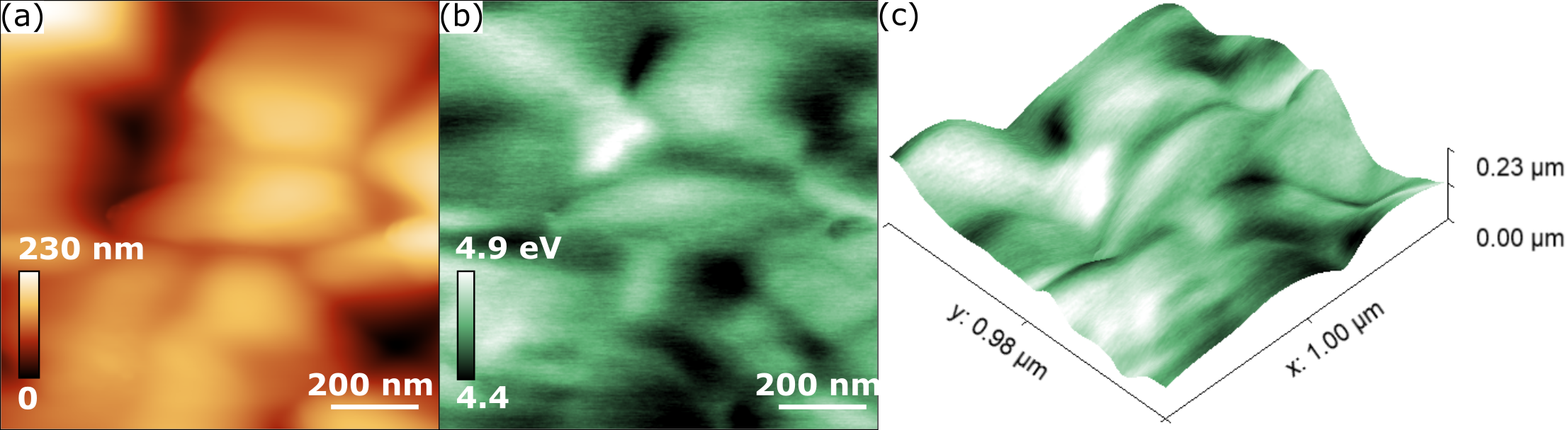}
    \caption{FM-KPFM measurement on a non- air exposed polycrystalline CIGSe sample. 
    (a) Topography map showing grains and GBs, (b) workfunction map showing up to 500\,meV peak to peak variations. 
    (c) 3D representation of the topography overlaid with the workfunction map.}
    \label{F:6}
\end{figure}

Finally, FM-KPFM measurements were also performed on polycrystalline CIGSe absorbers. The sample was grown by multi-stage co-evaporation and transferred into the SPM system via an UHV suitcase. 
The sample did not see background pressures higher than 10$^{-8}$\,mbar during the transfer. 
\textbf{Fig.~\ref{F:6}\,(a,b)} depict the topography and the KPFM measurements acquired on that absorber. 
In the topographic image, well-defined grain boundaries could be observed with valley to peak height differences of approximately 200\,nm. 
The workfunction map showed large inhomogenities up to 500\,meV. 
The average workfunction is close to the one of the non-air-exposed epitaxial CISe surface, which is very reasonable. 
From the separate topography and workfunction maps, it is hard to differentiate between workfunction and facet contrast. 
In \textbf{Fig.~\ref{F:6}\,(c)} a three-dimensional overlay of the topography and the CPD map is shown. It becomes clear that again most of the contrast is related to different parts of the grains, i.e. facets in analogy to the MAPI samples presented in this study. 
Therefore, 20 line profiles perpendicular to the grain boundaries are presented in \textbf{Fig.~\ref{FS:12}}\,(retrace scan) and \textbf{Fig.~\ref{FS:13}}\,(trace scan).
The line profiles show that the vast majority of the contrast that can be observed in the workfunction maps arise from the facets of the sample surface (i.e. the positions of the the minima of the GBs do not coincide with minima/maxima in the workfunction). 
We conclude that FM-KPFM measurements under UHV conditions for the non-air-exposed CIGSe sample showed no measurable contrast at most of the GBs. However, we note that due to the large facet related workfunction contrast, small CPD changes at the GBs of a few mV cannot be ruled out. This result is astonishing since it suggests that in high performance CIGSe GBs do not exhibit relevant band bending. Additional measurements on post-deposition treated CIGSe samples are mandatory in order to understand the impact on the electronic landscape of the CIGSe GBs under conditions where the surface is not exposed to air.

    \section{Conclusions}

In this work we have discussed in detail how rough polycrystalline thin films impact the accuracy of the KPFM results. 
In AM-KPFM, the dominating long-range electrostatic force between the surface and the cantilever leads to a strong topographic cross-talk, which results in erroneous values of the grain boundary potentials. 
This is corroborated by electrostatic calculations, which allow to reproduce the measured CPD maps. 
The results unambiguously show that the GB potential values measured with AM-KPFM in lift-mode, as often observed in literature, are not correct.
 
In contrast, the cantilever-induced cross-talk in FM-KPFM can be neglected, which is a direct consequence of the shorter range of the electrostatic force gradient compared to the electrostatic force itself. 
Measurements on high quality non air exposed CIGSe absorbers show that the work function variations at the surface are dominated by facet related work function variations and GBs do not exhibit a measurable contrast. This large facet related contrast of up to 500\,meV is in agreement with a reports of CuGaSe$_2$ grown on ZnSe \cite{Sadewasser2002High-resolutionSurfaces} and some reports carried out on halide perovskites \cite{Leblebici2016,Gallet2019Fermi-levelPerovskites}.

Halide perovskites show either a facet-related contrast (this work), or a weak GB contrast with values smaller than 100\,meV. This contrast depends sensitively on the surface state (annealed versus as-grown \cite{Gallet2021Co-evaporationStability}) or the type of surface passivation \cite{Krishna2021NanoscalePhotovoltaics}. 

In agreement with the measurements on polycrystralline absorbers, epitaxial non-air-exposed CISe samples also show prominent facet contrast, originating from (112) and (312) surface terminations. 
Prolonged exposure to the residual gas inside the UHV chamber or air exposure resulted in a complete disappearance of this contrast. We showed that a mild UHV annealing at 200$^\circ$C is an effective tool to recover the pristine properties of the absorber layer. 
On the other hand, annealing of air-exposed samples resulted in a different workfunction distribution, which we related to the formation of oxides on the surface of the absorber. 
These results corroborate that air exposure needs to be circumvented and cannot be compensated by UHV annealing.

Future studies should focus on understanding how these strongly varying surface dipoles, that are responsible for the facet-related contrast  can be modified by post deposition treatments to form an interface with low number of defects. Ultimately, the impact of these workfunction variations on device performance needs to be understood.

    \section*{Conflict of Interest}

The authors declare that they have no known competing financial interests or personal relationships that could have appeared to influence the work reported in this paper.

    \section*{Acknowledgments}

The authors acknowledge financial support from the Luxembourg Fonds National de la Recherche (“SUNSPOT”, Nr. 11244141; GRISC, Nr. 11696002; and SeVac Nr. C17/MS/11655733). The authors acknowledge Dr. Florian Ehre for processing the polycrystalline CIGSe sample, Dr. Michele Melchirorre for the EDX/SEM data, and the technicians of the SPM laboratory, Dr. Bernd Uder, Dr. Ulrich Siegel and Nicolas Tournier for technical support. E.M.L. acknowledges Dr. Carlos Costa from Nanotechnology National Laboratory – LNNano/CNPEM under the research proposal AFM-25904.

    \section*{Appendix A. Supplementary material}

The SI shows complementary KPFM data and simulations.

\bibliographystyle{elsarticle-num-names}
\bibliography{references}

\clearpage

\appendix
    \maketitle
\section{Supplementary material}

\newcommand{\hbAppendixPrefix}{S}
\renewcommand{\thefigure}{\hbAppendixPrefix\arabic{figure}}
\setcounter{figure}{0}

\renewcommand{\thetable}{\hbAppendixPrefix\arabic{table}}
\setcounter{table}{0}

\renewcommand{\theequation}{\hbAppendixPrefix\arabic{equation}}
\setcounter{equation}{0}

As shortly described in the experimental section of the main manuscript, KPFM is a powerful technique to measure the workfunction of a sample with nanometer resolution. The relation between the workfunction and the CPD value measured via KPFM is defined as:
\begin{equation}
    CPD = \frac{(\Phi_{s} - \Phi_{p})}{e}
    \label{E:1}
\end{equation}
where $\Phi_s$ is the workfunction of the sample, $\Phi_p$ the workfunction of the probe and $e$ is the elementary charge.
It is therefore, fundamental to know if the V$_{DC}$ bias is applied to the probe or to the sample because it inverts the CPD sign measured in the KPFM. 
To avoid any misunderstanding and to facilitate the comparison between different setups the DC voltage in this manuscript is always applied to be a sample bias. Consequently, the results displayed in \textbf{Figure\,\ref{S:1}(a)} of the main manuscript are all converted such that the bias is assumed to be applied to the sample  (see \textbf{Table \ref{TS:1}}).
\begin{table}[H]
    \caption{Summary of the references used in \textbf{Figure \ref{F:1} (a)}. References with no information regarding the environment were assumed to be performed under  ambient conditions.}
    \centering\includegraphics[width=1.0\linewidth]{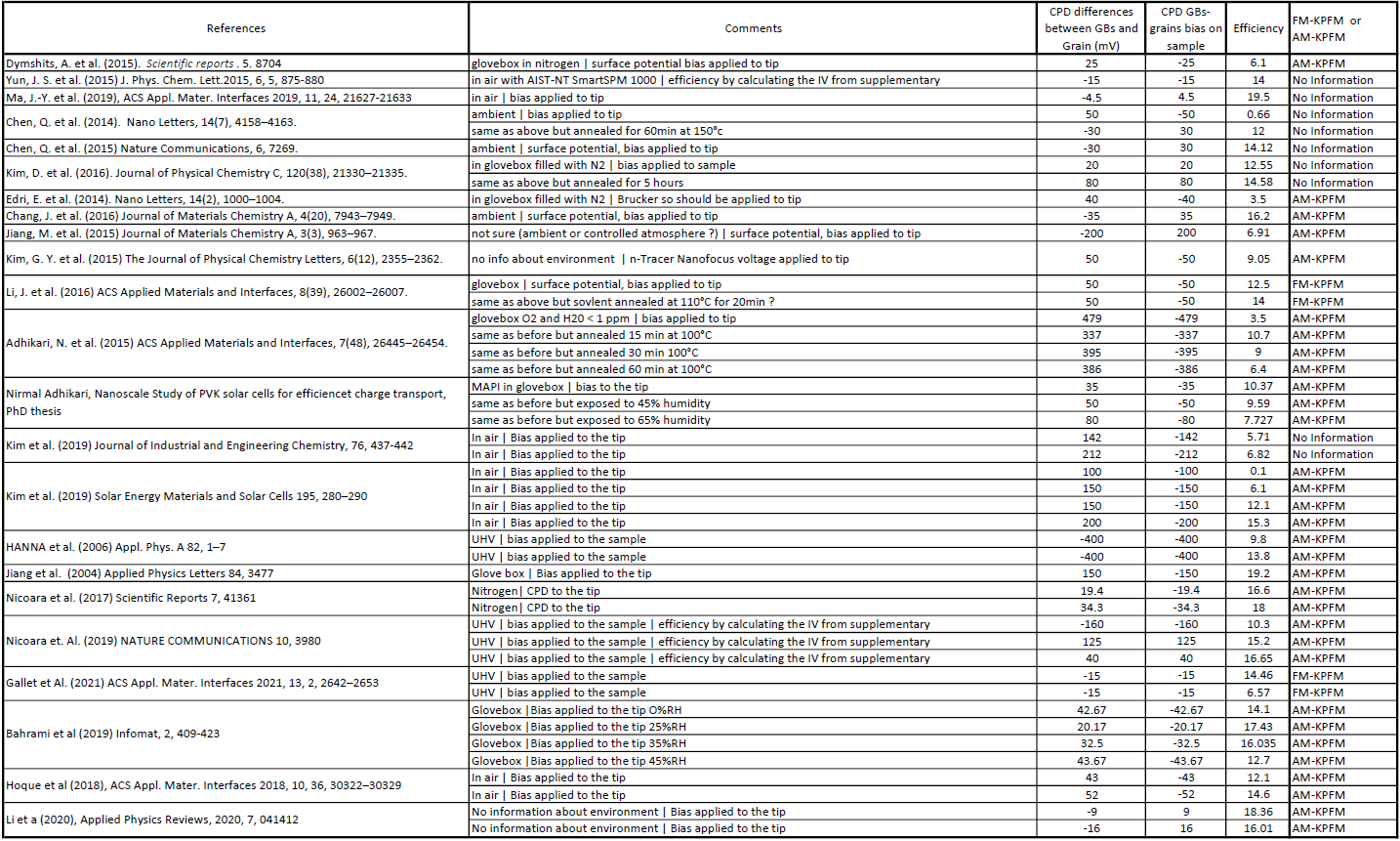}
    \label{TS:1}
\end{table}

\textbf{\subsection{Differences between FM- and AM-KPFM}}

The basic principle of KPFM is to measure the sample surface potential by probing the electrostatic forces between the sharp conductive SPM probe and the sample. 
For this, an AC voltage applied between the probe and the sample surface will generate an oscillating electrostatic force whenever there is a CPD between them.
A feedback loop then applies a DC bias (V$_{DC}$) between the probe and the sample to nullify this oscillating electrostatic force. 
When the DC bias is equal to the unknown sample surface potential, the oscillating electrostatic force will be null, and the KPFM electronics will record the applied voltage to build a surface potential map of the sample. 

In AM-KPFM, changes in the amplitude of the oscillating electrostatic force are directly measured via the probe deflection at a determined frequency. The KPFM feedback loop applies a V$_{DC}$ bias in order to nullify the amplitude oscillation. 
This procedure can be done in a single-pass mode by using phase-sensitive detection of the mechanical oscillation at a frequency $f_e$ distinct from the probe fundamental resonance $f_0$ \citeNew{nonnenmacher_kelvin_1991}.
Alternatively, it can also be done in double-pass mode, where the topography is measured in a first step, then the probe is retracted and kept a few nanometers away from the sample surface to record the potential information \citeNew{Jacobs1997SurfaceSPM}. In double-pass mode, the electrostatic force is decoupled physically from the short-range forces. 

Conversely, in FM-KPFM, a resonance frequency shift ($\Delta{f}$) is induced by the V$_{AC}$ oscillating voltage. In this case, the fundamental mechanical resonance ($f_0$) of the probe is modulated by a frequency $f_{mod}$ (1 - 2 kHz), generating sidebands at $f_0 \pm f_{mod}$, which are proportional to the gradient of the electrical forces \citeNew{Colchero2001ResolutionMicroscopy}. 
The KPFM feedback loop then applies a V$_{DC}$ bias to nullify the frequency shift at $f_0 + f_{mod}$.

In principle, AM-KPFM and FM-KPFM should give exactly the same surface potential values, however, it has been demonstrated in the literature that this is not the case \citeNew{Glatzel2003AmplitudeMicroscopy, Zerweck2005AccuracyMicroscopy, Diesinger2012CapacitiveKPFM}. 
The differences are generally attributed to the higher lateral sensitivity of the FM-KPFM, in which the capacitive coupling between the cantilever and the sample can be avoided. 
Due to the simpler technical implementation, AM-KPFM is the most frequently used setup under ambient conditions, however, FM-KPFM provides more consistent results \citeNew{Axt2018KnowDevices}. FM-KPFM is also the most convenient technique under UHV conditions because the distance between the probe apex and the sample surface can be better controlled.

\begin{figure}[H]
    \centering
    \includegraphics[width=0.8\linewidth]{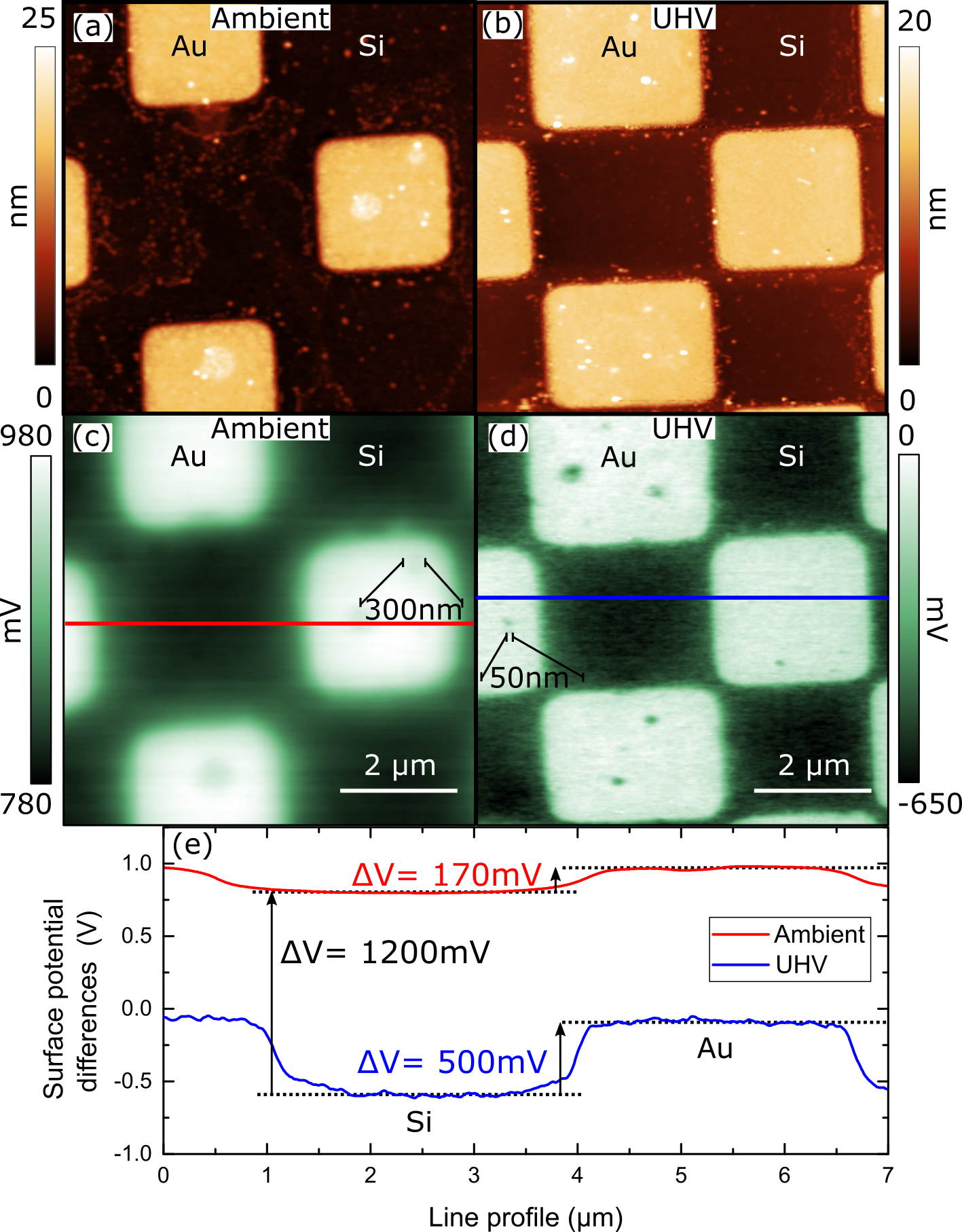}
    \caption{Topography and KPFM measurements of an Au patterned  silicon substrate measured with AM-KPFM  under ambient conditions (a,c) and  with FM-KPFM under ultra-high vacuum conditions (b,d). The bright squares in the topography images represent the Au regions. 
    (e) extracted topography line profiles under ambient (red) and UHV conditions (blue). 
    The Au structures are around 15\,nm high. Under ambient conditions, the surface potential on Si is 1200\,mV higher than under UHV. 
    The surface potential difference between gold and silicon observed under UHV is similar to to the reported workfunction difference of Au and Si.}
    \label{FS:1}
\end{figure}

To confirm the results from the literature and to make sure that our setups are working properly, we measured a commercially available KPFM reference sample composed of a gold pattern on silicon substrate.
\textbf{Figure \ref{FS:1}} shows a KPFM measurement on a gold patterned silicon substrate (Anfatec Instruments AG) measured under ambient conditions in AM mode (\textbf{Figure \ref{FS:1}} (a, c)) and under UHV conditions in FM-mode (\textbf{Figure \ref{FS:1}} (b, d)). 
In the topography images (\textbf{Figure \ref{FS:1}} (a, b)), the bright regions of the images represent the gold patches. 
The equivalent surface potential maps are depicted in the \textbf{Figure \ref{FS:1}}(c, d), where the gold patches present higher workfunction values, in accordance to the expected tabulated workfunction values for both materials (5.3\,eV for gold and 4.85\,eV for Si with SiO$_2$ native layer). 

Qualitatively, the measured contrast in FM-KPFM under UHV and AM-KPFM under ambient conditions were very similar; however, the lateral resolution and the exact values of the workfunction were different. 
Structures down to 50\,nm could be resolved in the FM-KPFM under UHV, which are close to the probe radius (25\,nm) and the pixel resolution for this image (23.3\,nm). 
On the other hand, we could not distinguish structures smaller than 300\,nm in the AM-KPFM under ambient conditions even with the same probe radius and image resolution. 
\textbf{Figure \ref{FS:1}}(e) depicts the line profiles of the surface potential extracted from the images under ambient and UHV conditions, red and blue lines, respectively. 
We observed a shift in the surface potential of around 1200\,mV between the two line-profiles that was not expected since we used the same Pt-Ir probe. 
Additionally, we could see that the contact potential difference between the gold and silicon under ambient conditions was reduced by a factor of three compared to UHV conditions (170\,mV compared to 500\,mV). 
We attributed these well-known effects to the presence of a water layer on the sample surface exposed to air that adds a dielectric contribution and stray capacitance field to the value measured by KPFM \citeNew{Salomao2015DeterminationMicroscopy, burgo_electric_2011, Sugimura2002PotentialMicroscopy}.

The AM-KPFM under ambient conditions is sensitive to the electrical force, and consequently, the full cantilever contributes to the signal due to a capacitance coupling with the sample. 
FM-KPFM under UHV is sensitive to the gradient of the electrical force, which is more confined to the probe apex and, therefore, reduces the cross-talk and improves the resolution. 
From these measurements, we concluded that AM-KPFM under ambient conditions as well as FM-KPFM under UHV conditions could resolve the workfunction difference between the gold/Si pattern, however, as already shown in the literature, FM-KPFM yielded a better lateral resolution and values of the workfunction difference closer to the expected one ($\approx$\,0.5eV) \citeNew{Axt2018KnowDevices, Glatzel2003AmplitudeMicroscopy, Zerweck2005AccuracyMicroscopy}.

\begin{figure}[H]
    \centering
    \includegraphics[width=0.7\linewidth]{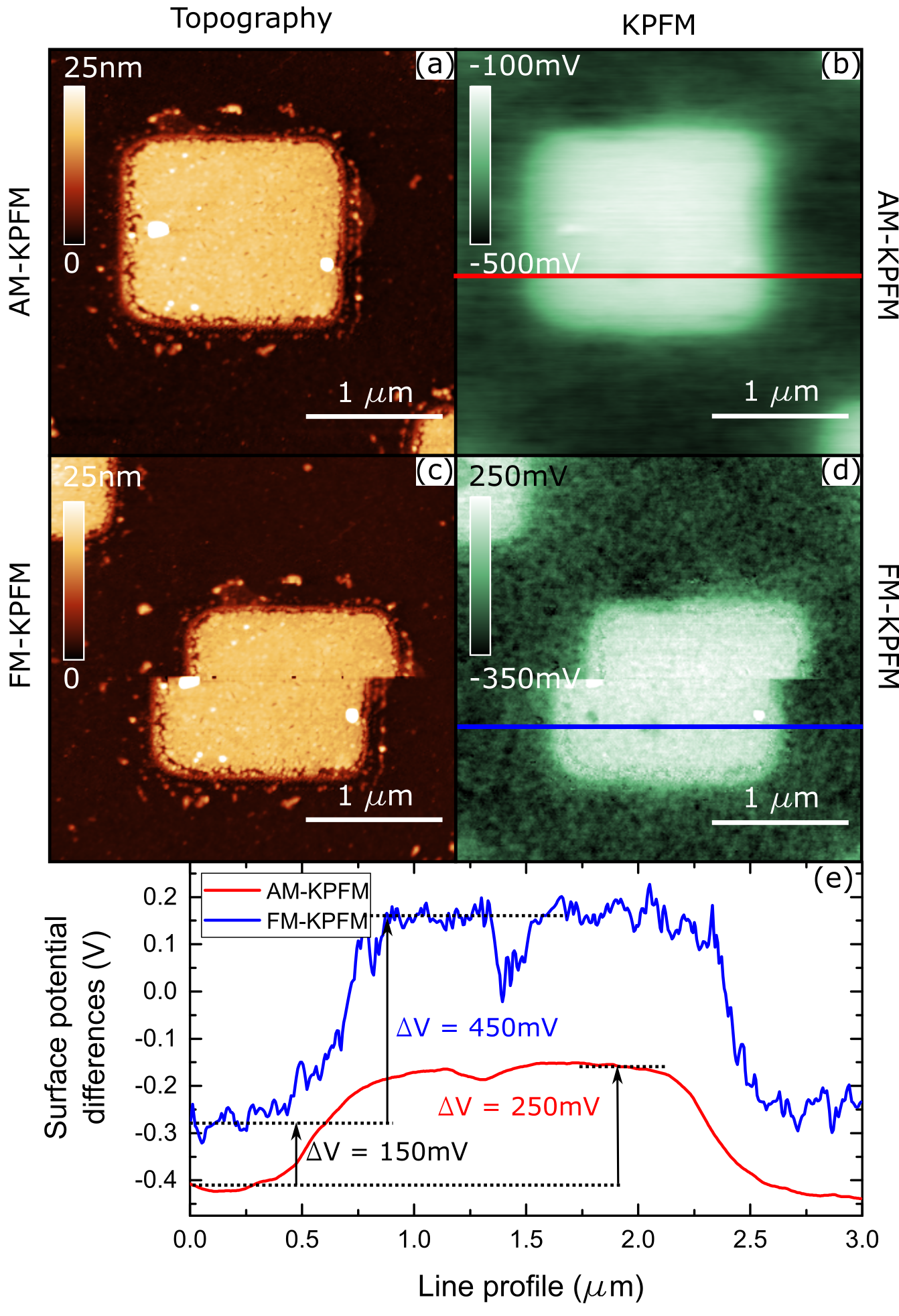}
    \caption{AM-KPFM and FM-KPFM acquired on the same spot under UHV conditions. 
    From topography (a) and (c), we can identify the same spot of the sample even with a \textit{probe jump} in the FM-KPFM image. 
    Potential images (b) and (d) clearly show the superior resolution of the FM-KPFM mode. 
    (e) Surface potential line profiles extracted from the AM-KPFM and FM-KPFM, respectively red and blue lines. 
    We observed a 150\,mV difference between both techniques on the Si surface. 
    When comparing the surface potential differences between gold and silicon we do measure a difference of about 200\,mV. This value is smaller than shown in \textbf{Figure \ref{FS:1}} where the difference between both techniques was 330\,mV presumably due to an additional water layer on top of the sample surface.}
    \label{FS:2}
\end{figure}

To corroborate that FM-KPFM provides better quantitative workfunction measurement than AM-KPFM independently from the environment, we used both methods under UHV and at the same position of the gold pattern sample. These measurements are depicted in \textbf{Figure \ref{FS:2}}. 
Similarly to the previous images, the brighter regions in the topography (\textbf{Figure \ref{FS:2}} (a, c)) represent the gold patches, which show higher values for the workfunction (\textbf{Figure \ref{FS:2}} (b, d)). 
The measured CPD difference between gold and Si in AM-KPFM was around 50\% smaller than the one measured by FM-KPFM.
One can also see from the images and line profiles that FM-KPFM provides much more lateral information than AM-KPFM.
From \textbf{Figures \ref{FS:1} and \ref{FS:2}}, we can state that ambient conditions massively affect the KPFM results and that FM-KPFM offers better lateral resolution combined with a more precise values for the workfunction measurements. 
The measurements in UHV allowed us to conclude that, independent of the environment, AM- and FM-KPFM measurements did not yield the same result on exactly the same spot of the same sample.

\textbf{\subsection{Individual forces that act on an AFM probe}}

\begin{figure}[H]
    \centering
    \includegraphics[width=0.8\linewidth]{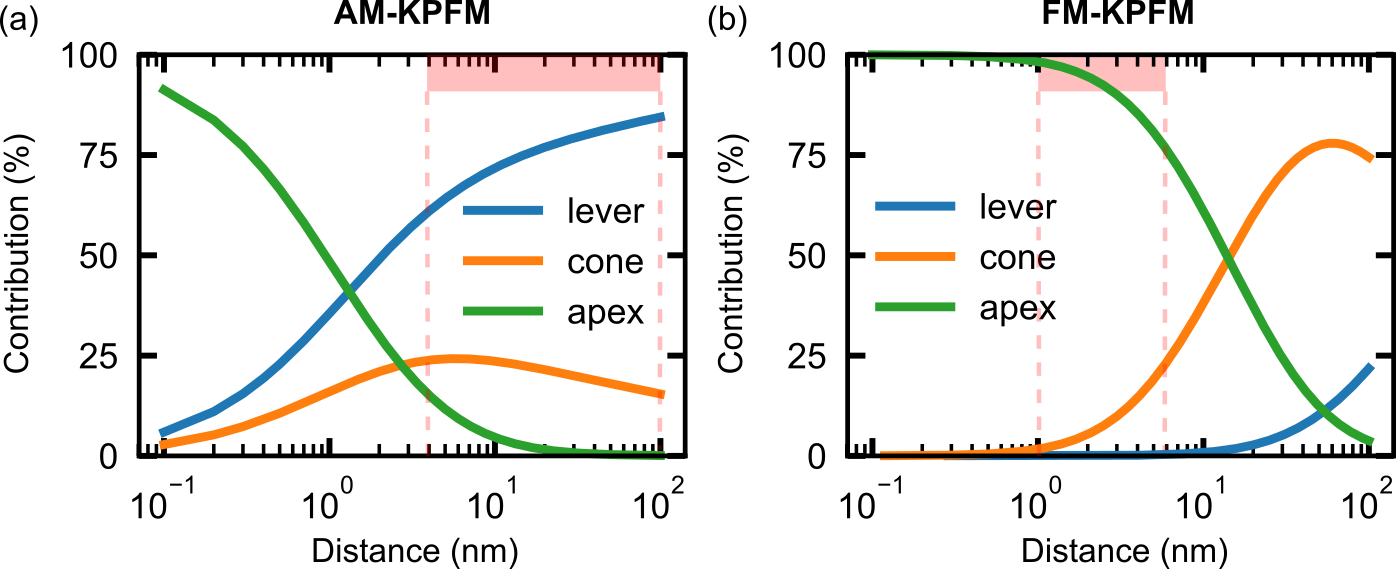}
    \caption{Force and force gradient contributions to the KPFM signal (a) AM-KPFM and (b) FM-KPFM for each part of the probe.
    Blue, orange and green represent the cantilever, the cone and the probe apex, respectively.
    In each graph, the pink region marks  the typical working distance  for AM-KPFM (double pass) and FM-KPFM. Calculations were done according to {\cite{Wagner2015KelvinDevices}} without considering the probe oscillation.}
    \label{FS:3}
\end{figure}

In analogy to  \citetNew{Wagner2015KelvinDevices} we defined the probe as  being composed of three parts, cantilever, cone and probe apex. These parts are marked in \textbf{Figure \ref{F:1} (b)}.
It is essential to understand from where the dominating force to the  KPFM signal arises from. FM-KPFM is sensitive to the gradient of the electrostatic force and  AM-KPFM is sensitive to the electrostatic force \citeNew{Glatzel2003AmplitudeMicroscopy, Zerweck2005AccuracyMicroscopy, Jacobs1997SurfaceSPM}. We calculated the individual contributions without considering the oscillation of the probe due to the tapping mode. \textbf{Figure \ref{FS:3}} depicts the contribution of   the cantilever, the cone and the probe apex as a function of the probe sample distance for the AM and FM case. The dashed region highlights typical probe sample distances used experimentally. In our case, a Python script was written using the \textbf{equations \ref{eq:s2} \ref{eq:s3} and \ref{eq:s4}}.

\begin{equation}
    F_{lever}=-\frac{1}{2}\frac{\epsilon_0 A}{(z+H)^2}U_{ts}^2
    \label{eq:s2}
\end{equation}

\begin{equation}
    F_{cone}=-\pi \epsilon_0 k^2 U_{ts}^2 \left[ \ln\frac{H}{z+\tilde{R}} -1 + \frac{R \cos^2{\theta_0} / \sin{\theta_0}}{z+\tilde{R}} \right]
    \label{eq:s3}
\end{equation}

\begin{equation}
    F_{apex}=-\pi \epsilon_0 R U_{ts}^2 \left(\frac{1}{z} - \frac{1}{z+\tilde{R}}\right)
    \label{eq:s4}
\end{equation}

Where $k^2 = (\ln\tan{\theta_0/2})^{-2}$, $\tilde{R} = R(1-\sin{\theta_0}$, R is the apex radius of the probe, $\theta_0$ is the half open of the cone.

The marked regions in the graphs show the working distance range for each KPFM mode, from that we can conclude that the main contribution to the KPFM signal in AM-KPFM double-pass mode is coming from the cantilever, while in FM-KPFM it comes from the probe apex.


\textbf{\subsection{Additional figures}}

\begin{figure}[H]
    \centering
    \includegraphics[width=0.8\linewidth]{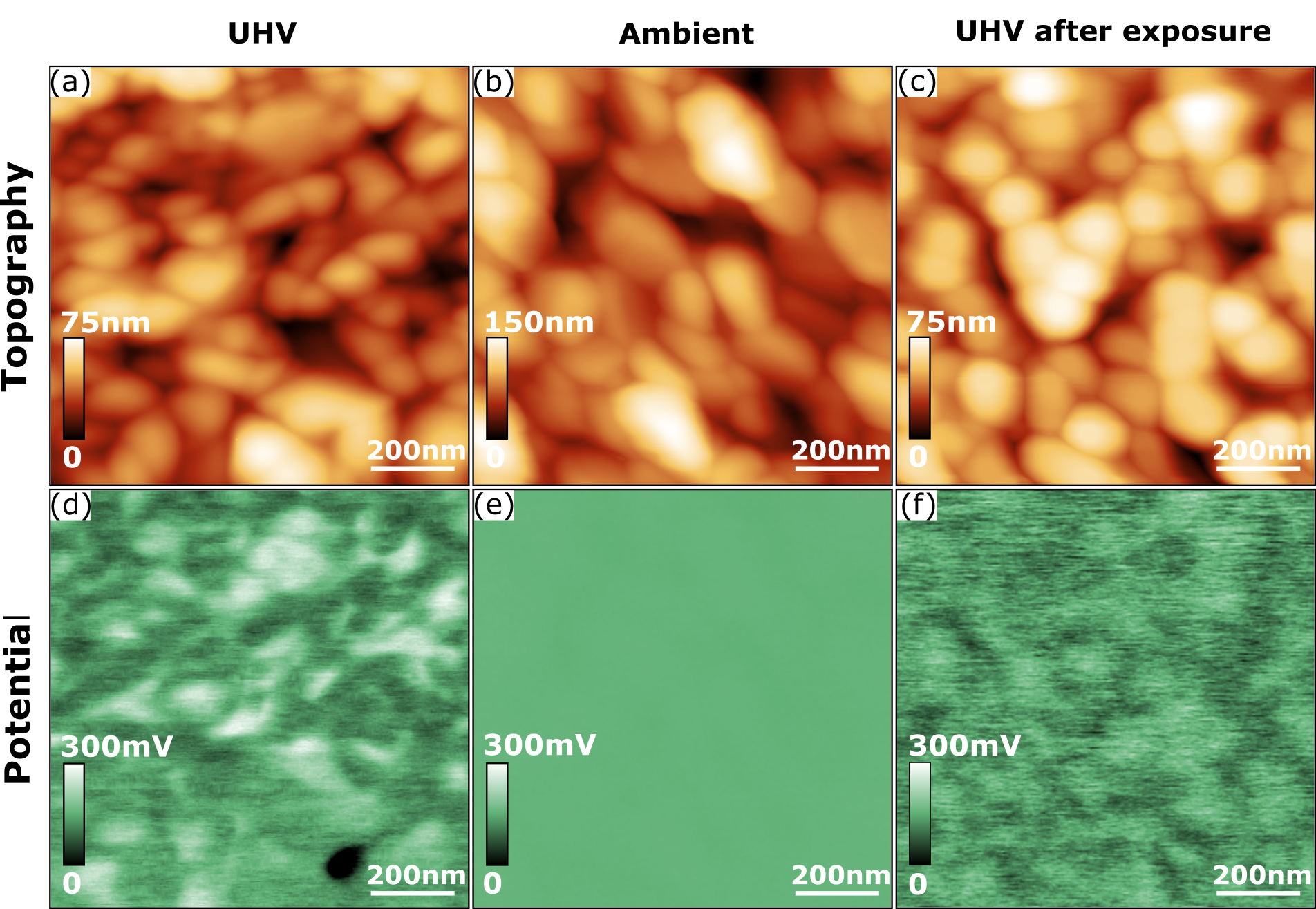}
    \caption{Complement to \textbf{Figure \ref{F:2}}. Topography (a-c) with the respective surface potential (d-f) measured under UHV (FM-KPFM), ambient (AM-KPFM) and UHV after exposing it to air (FM-KPFM). (d) The surface potential measured under UHV without previous exposure to air shows a facet dependent contrast.  (e) AM-KPFM under ambient conditions. Almost no signal is observed when compared to the results measured in UHV. (f) Stronger (none facet-related contrast)  for the air exposed sample measured in UHV. This sample was kept for 17 days under nitrogen atmosphere before introducing it to UHV and it was air exposed for roughly 2 days before the second measurement under UHV.}
    \label{FS:4}
\end{figure}

\begin{figure}[H]
    \centering
    \includegraphics[width=0.8\linewidth]{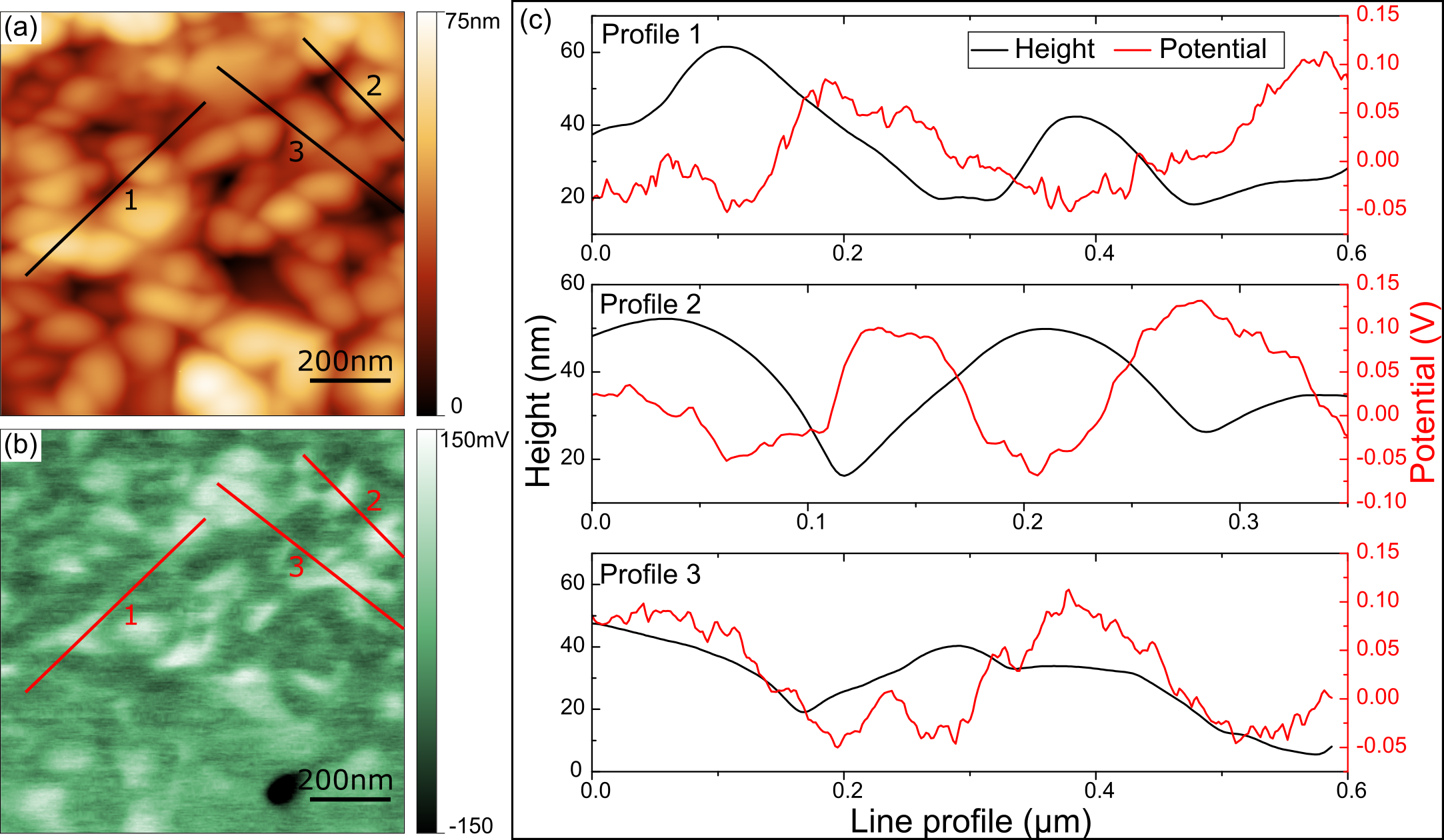}
    \caption{Facet related surface potential. (a,b) show the topography and the surface potential image for the MAPI absorber. 
    Line profiles, labeled as 1, 2 and 3 are plotted in (c). Black and red line profiles represent topography and potential data respectively.
    From the line profiles we do see that changes in the workfunction are related to the different surface facets of the crystal. 
    Negligible or no workfunction contrast was observed at the GBs}
    \label{FS:6}
\end{figure}

\begin{figure}[H]
    \centering
    \includegraphics[width=0.8\linewidth]{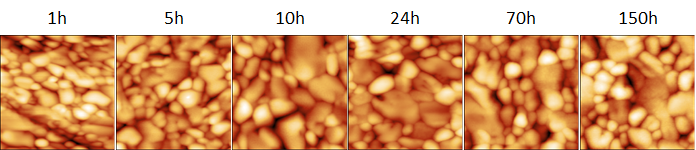}
    \caption{Temporal variations of the topography of a MAPI absorber under ambient conditions. No intentional change of the probe or sample position during the experiment. The strong topography variations over time are due to an enhanced thermal drift combined with decomposition of the sample surface. The large fluctuations of temperature and humidity inside the room over 150 hours were presumably responsible for the changes in perovskite morphology.}
    \label{FS:7}
\end{figure}

\begin{figure}[H]
    \centering
    \includegraphics[width=0.8\linewidth]{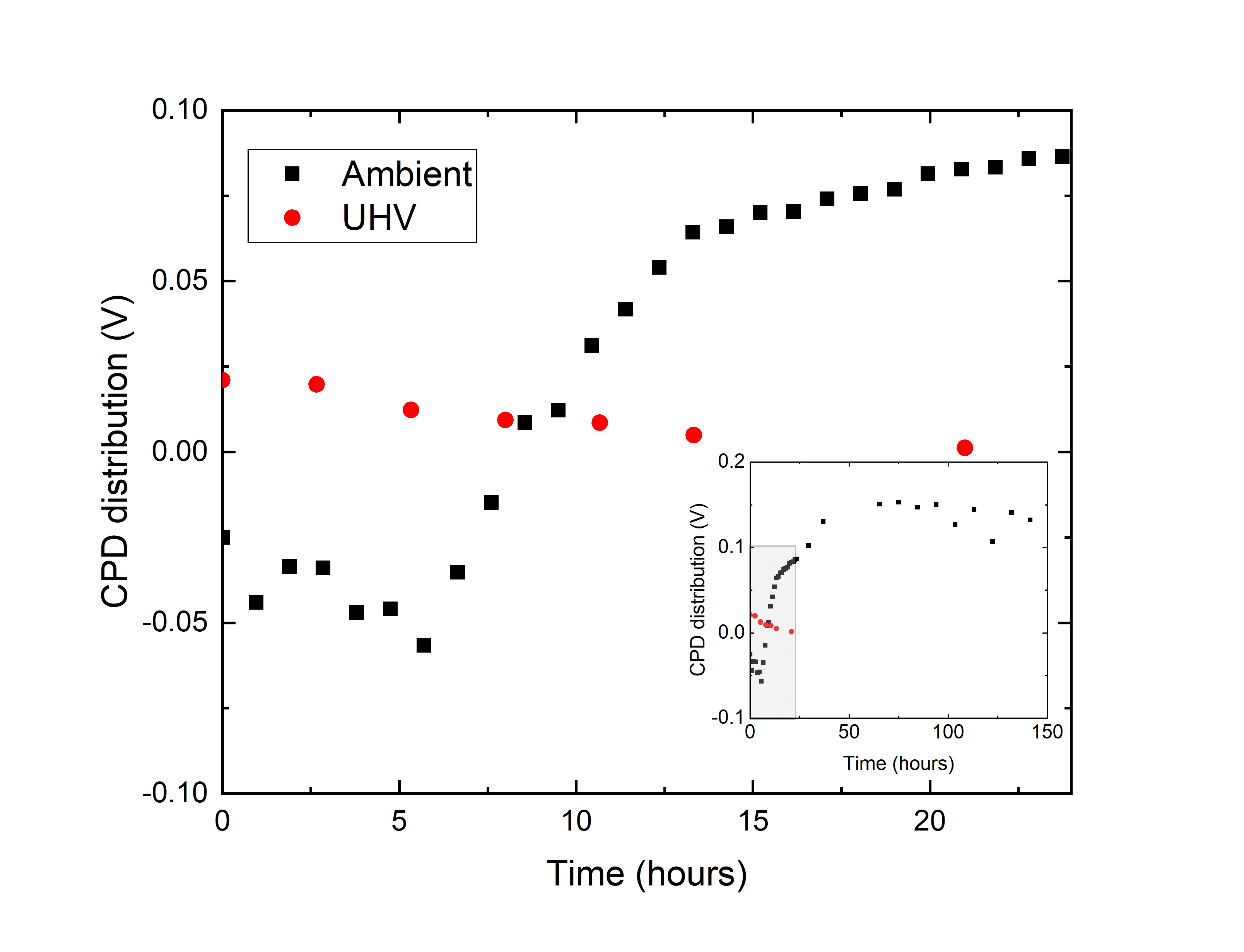}
    \caption{Changes in average CPD as a function of the scanning time extracted from AM-KPFM maps measured on perovskite absorbers under ambient conditions (black squares) and FM-KPFM maps in UHV  (red dots). For FM-KPFM case in UHV  less than 25\,mV CPD variations were observed. Under ambient conditions more than 150\,mV changes were observed within the same time scale, and around 200\,mV over 150\,hours (see inset).}
    \label{FS:8}
\end{figure}

\begin{figure}[H]
    \centering
    \includegraphics[width=0.8\linewidth]{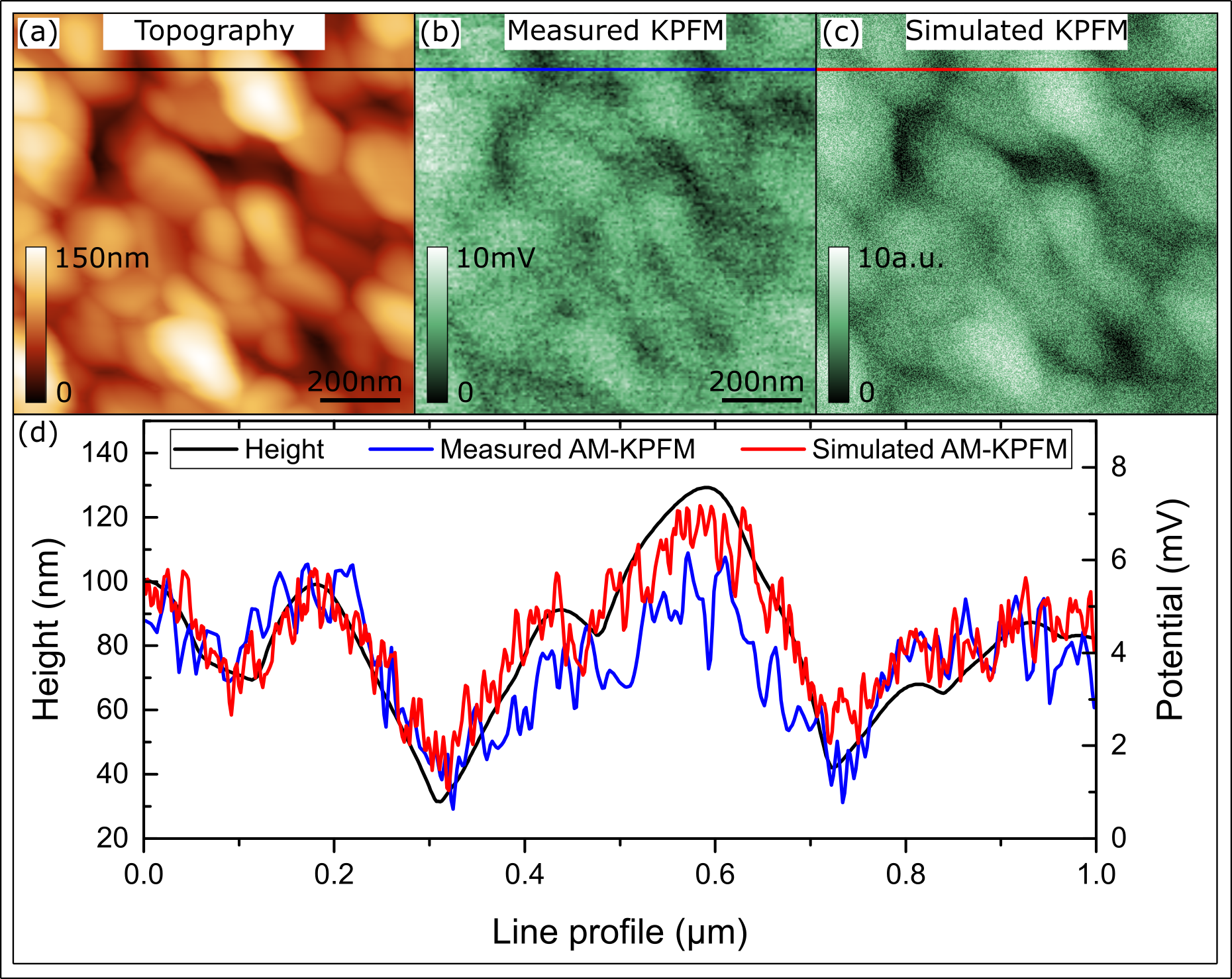}
    \caption{AM-KPFM performed on a MAPI absorber under ambient conditions from the same sample as shown in \textbf{Figure \ref{F:2}}. (a,b) measured topography and surface potential images. (c) simulated AM-KPFM. (d) line profiles from topography, measured and simulated AM-KPFM. The strong correlation between the measured and the simulated data suggests that the majority of the AM-KPFM measured signal originates from a cross-talk with the topography, in analogy to the one observed in \textbf{Figure \ref{F:3}}.}
    \label{FS:5}
\end{figure}

\begin{figure}[H]
    \centering
    \includegraphics[width=0.8\linewidth]{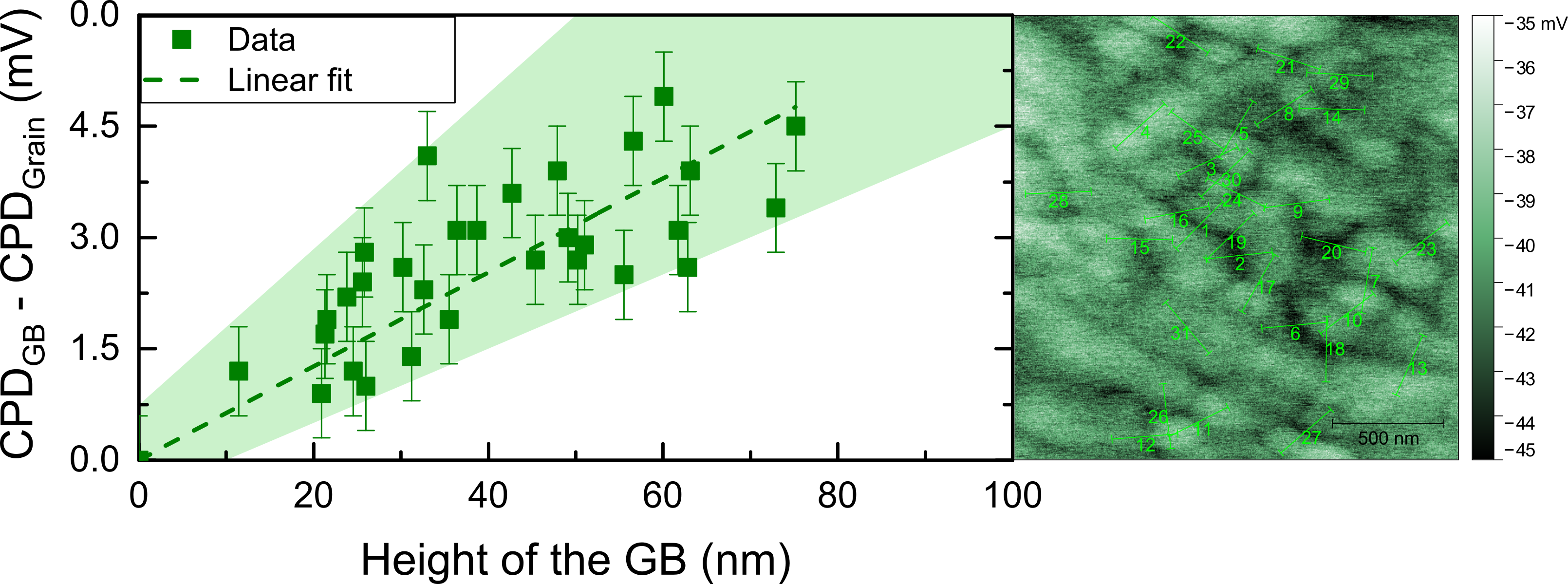}
    \caption{(a) CPD$_{GB}$ – CPD$_{Grain}$ versus the depth of the grain boundaries for the sample MAPI after 1\,hour exposed to air. Each dot in the graph represents the depth of the GB extracted from the 30 line profiles and the equivalent CPD differences extracted from (b). The highlighted green area depicts the envelope from the entire data-set. The plot shows a correlation between topography and CPD values, which cannot be physically explained as a charge accumulation.}
    \label{FS:9}
\end{figure}

\begin{figure}[H]
    \centering
    \includegraphics[width=0.8\linewidth]{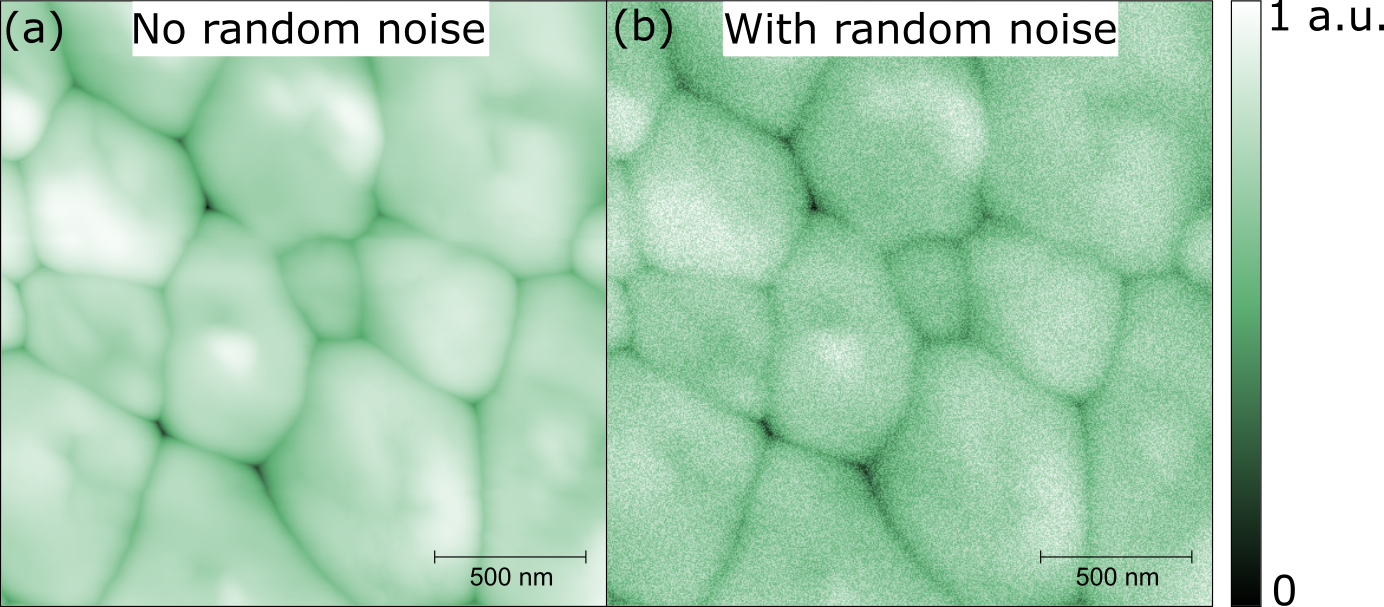}
    \caption{Random noise for simulated data. (a) Shows the simulated data without considering any instrument noise. 
    (b) Shows the same simulated data with a random noise of 2\,mV.}
    \label{FS:10}
\end{figure}  

\begin{figure}[H]
    \centering
    \includegraphics[width=0.8\linewidth]{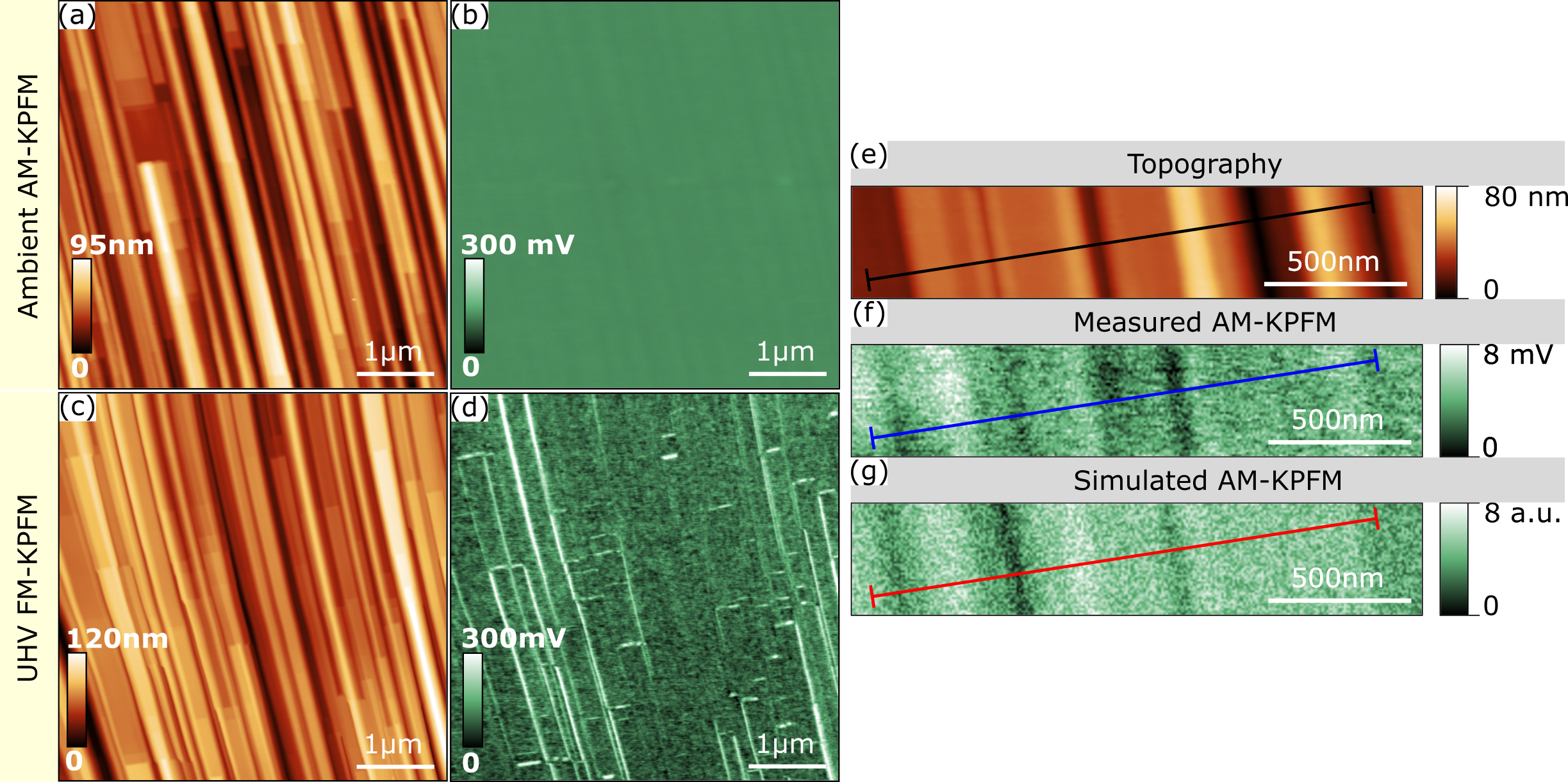}
    \caption{Epitaxial CIGS absorber topography and KPFM data already presented in  \textbf{Figure \ref{F:4}} with same scale for the surface potential. 
    (a,b) Depict topography and potential images for the AM-KPFM under ambient conditions. 
    Scale bar of the potential data was adjusted from 0 to 300\,mV for a better comparison with (d) FM-KPFM under UHV.
    (c-d) Depict topography and potential images for the FM-KPFM under UHV. 
    (e) Topography, (f) measured AM-KPFM and (g) simulated AM-KPFM for the epi-CISe. 
    The topography was used as input for the height change $\Delta h$ in \textbf{equation \ref{eq3}.} 
    }
    \label{FS:11}
\end{figure}

\begin{figure}[H]
    \centering
    \includegraphics[width=0.8\linewidth]{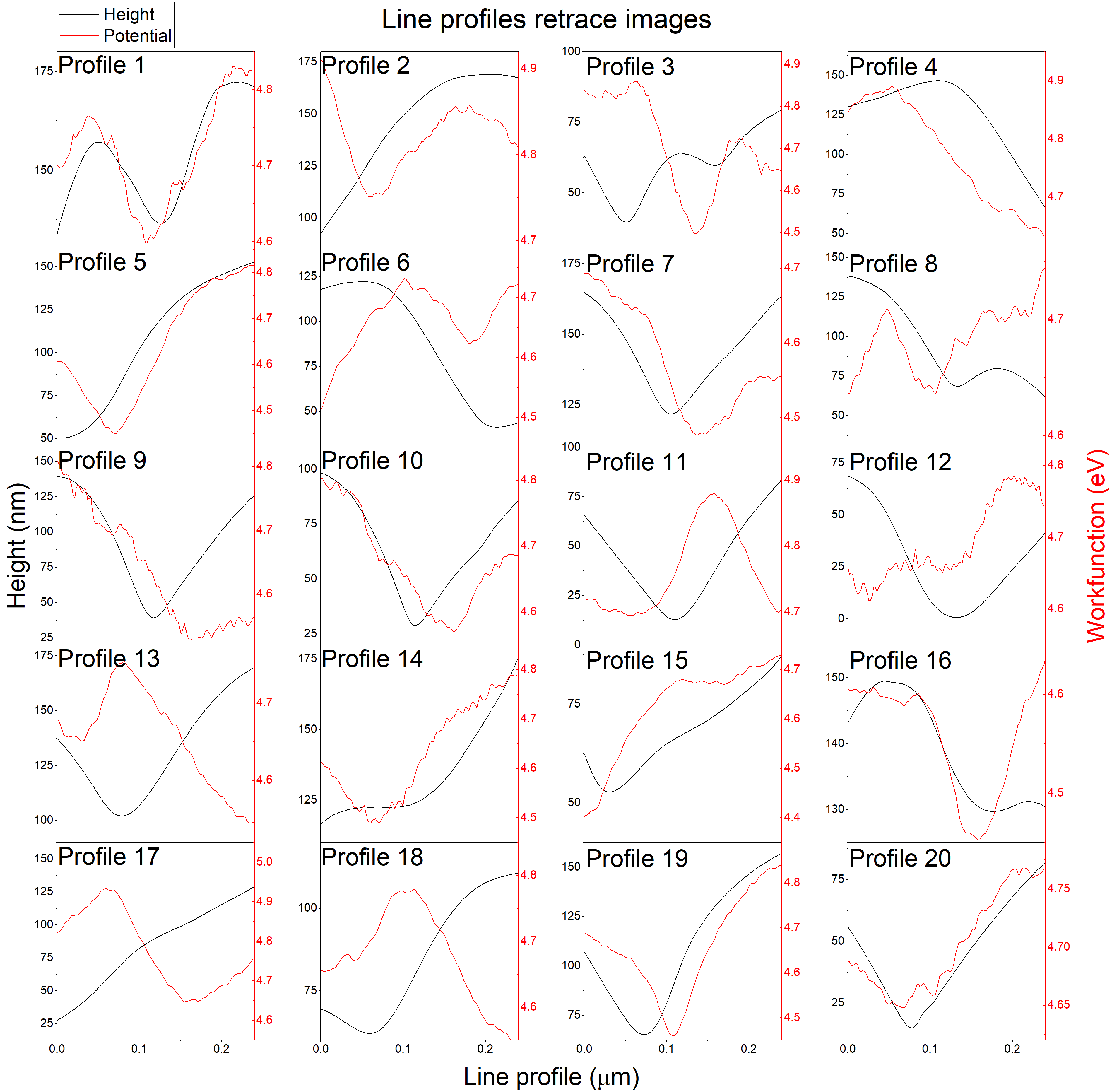}
    \caption{20 line profiles from retrace FM-KPFM images of the polycrystalline CIGSe sample depicted in \textbf{Figure \ref{F:6}}. Black solid lines represent the height and red solid lines represent the workfunction data.}
    \label{FS:12}
\end{figure}

\begin{figure}[H]
    \centering
    \includegraphics[width=0.8\linewidth]{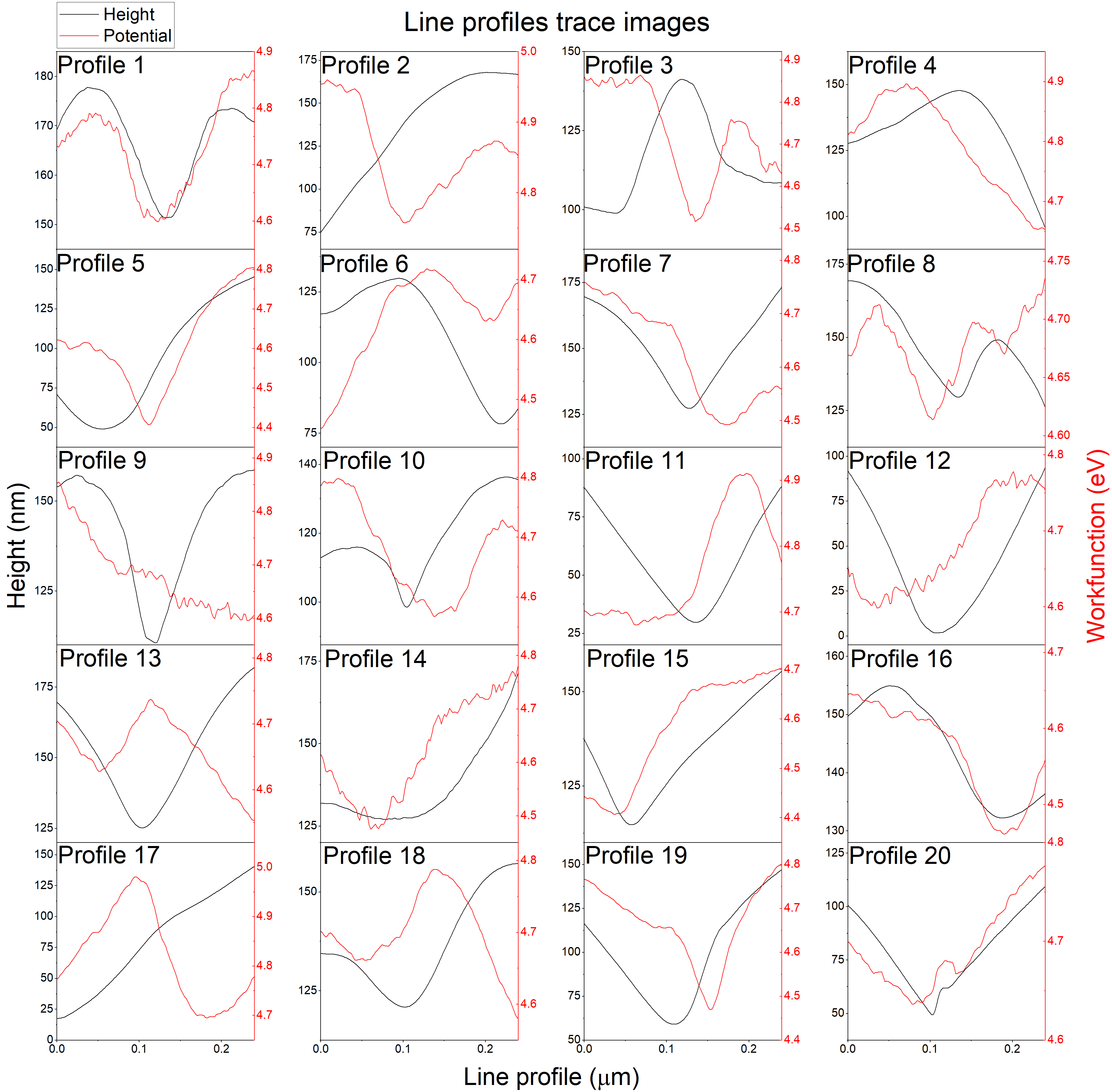}
    \caption{20 line profiles from trace FM-KPFM images of the polycrystalline CIGSe sample depicted in \textbf{Figure \ref{F:6}}. Black solid lines represent the height and red solid lines represent the workfunction data.}
    \label{FS:13}
\end{figure}

\bibliographystyleNew{elsarticle-num-names}


\end{document}